**Title**

# Transverse Load Optimisation in $Nb_3Sn$ CICC Design; Influence of Cabling, Void Fraction and Strand Stiffness

**Authors**

A. Nijhuis and Y. Ilyin

**Authors address**

University of Twente, Faculty of Science and Technology, Low Temperature Division, P.O. Box 217, 7500 AE Enschede, The Netherlands

**Short title**

Transverse load optimisation in $Nb_3Sn$ CICCs

**Abstract**

We have developed a model that describes the transverse load degradation in $Nb_3Sn$ CICCs, based on strand and cable properties, and that is capable to predict how such degradation can be prevented.

The $Nb_3Sn$ Cable In Conduit Conductors (CICC) for the International Thermonuclear Experimental Reactor (ITER) show a significant degradation in their performance with increasing electromagnetic load. Not only the differences in the thermal contraction of the composite materials affect the critical current and temperature margin, but mostly electromagnetic forces, cause significant transverse strand contact and bending strain in the $Nb_3Sn$ layers.

Here, we present the model for Transverse Electro-Magnetic Load Optimisation (TEMLOP) and report the first results of computations for ITER type of conductors, based on the measured properties of the internal tin strand used for the Toroidal Field Model Coil (TFMC). As input, the model uses data describing the behaviour of single strands under periodic bending and contact loads, measured with the TARSIS setup, enabling a discrimination in performance reduction per specific load and strand type.

The most important conclusion of the model computations is that the problem of the severe degradation of large CICCs can be drastically and straightforwardly improved by increasing the pitch length of subsequent cabling stages. It is for the first time that an increase of the pitches is proposed and no experimental data are available yet to confirm this beneficial outcome of the TEMLOP model. Larger pitch lengths will result in a more homogeneous distribution of the stresses and strains in the cable by significantly moderating the local peak stresses associated with the intermediate-length twist pitches. The twist pitch scheme of the present conductor layout turns out to be unfortunately close to a worst-case scenario.

The model also makes clear that strand bending is the dominant mechanism causing degradation. The transverse load on strand crossings and line contacts, abbreviated as contact load, can reach locally 90 MPa but this occurs in the low field area of the conductor and does not play a significant role in the observed critical current degradation. The model gives an accurate description for the mechanical response of the strands to a transverse load, from layer to layer in the cable, in agreement with mechanical experiments performed on cables.

It is possible to improve the ITER conductor design or the operation margin, by mainly a change in the cabling scheme. We also find that a lower cable void fraction and larger strand stiffness add to a further improvement of the conductor performance.

## 1. Introduction

The envisaged CICCs for the International Thermonuclear Experimental Reactor consist out of more than 1000 strands with a strand diameter of about 0.8 mm. The conductors have a void fraction (*vf*) of 33 % and are cabled by twisting in several stages, thus creating a wavy pattern of the strands throughout the cable. When cooling from reaction heat treatment to cryogenic operation temperature, the different thermal contraction between conduit material and strand bundle causes besides a contraction of the strands in axial direction, likely also bending of the strands. This bending strain is again moderated by the coil hoop stress during magnet operation. The conductors are carrying more than 50 kA in a magnetic field locally exceeding 13 T, hence subjecting them to severe transverse loading due to the Lorentz forces. This imposes distributed magnetic loads along each strand, but also cumulated loads from other strands transferred by the strand-to-strand contacts. These bending and contact loads on the $Nb_3Sn$ strands affect the critical current and creates a periodic strain variation along the filaments. The magnitude and periodicity of the periodic strain pattern in combination with the ability of a strand to redistribute the current between the filaments determines the impact on the critical current ($I_c$) and $n$-value[1,2].

Since the first tests of the Central Solenoid Model Coils (CSMC) in Japan, lots of effort has been spent to the understanding of the unexpected severe degradation of the conductors compared to that of the single strand performance [3-8]. Although many papers were published on this subject, they mostly jus describe the degradation giving possible explanations varying from sometimes-severe current non-uniformity to often-severe strand bending. Most of the explanations are supported by models but unfortunately, none of these models is able to provide an adequate solution to the problem although it was shown that a lower void fraction and shorter cabling pitches partly confines the degradation [3]. In the meantime, it proven experimentally that mechanical support of the strands improved the performance significantly [9] and the ITER conductor design has already been modified compared to the CSMC layout [10]. It is assumed that the resistance to the degradation is increased by using a steel jacket, providing thermal pre-compression in reducing the tensile strain levels being associated with strand bending [3,11]. The void fraction was reduced from 36 % to 33 % and the non-copper material in the cross section was increased by 25%. In addition, newly developed, so-called high $J_c$ advanced $Nb_3Sn$ strands are being pursued in attempt to compensate for the performance loss.

Nevertheless, today it seemed accepted that the transverse load inherently leads to a severe degradation in the transport properties due to strand deformation. Although not conclusively proven, some of the analysts in the fusion community suggest that about 30 % to 40 % of the $Nb_3Sn$ layer in the strand material of the cable is breaking [3,10]. This would implicitly suggest that the cable in conduit concept for the large ITER type of conductors is beyond the limits of efficient application. In view of this thought, the development of advanced high $J_c$ strands would not immidately lead to an appropriate solution either, if a significant fraction of the $Nb_3Sn$ material then breaks in the cable. It is evident that a method to solve this severe degradation would lead to significant cost savings when requiring less superconducting strand or a significant improvement of the ITER magnets performance in terms of operation margin.

We believe that we have revealed a straightforward and economical method to overcome a significant part of this excessive reduction in performance. A newly developed model describing the mechanical response of the individual strands within a cable bundle, clearly shows that applying longer cabling pitches provides a convenient solution against transverse load degradation. In all probability, insufficient practical experience with varying cabling pitches, explains why this factor remained hidden until now. Moreover, it was assumed that only shorter cabling pitches could lead to a better performance obviously by shortening the bending beam [3], which is indeed confirmed by our model but for various reasons can be considered an un-practical solution. The model calculations on the present cable design are in good agreement with the deformation experiments that were performed on full-size ITER conductors in the Twente Cryogenic Press, as well as with the $I_c$ degradation observed in the CSMC tests [5].

An first essential parameter (also deduced from the press measurements) turns out to be the maximum possible compression of the cable, i.e. the available space for bending and contact deformation of strands. In addition, the strand mechanical properties determined from tests with the “Test ARangement for Strain Influence on Strands” (TARSIS) form the second key input for the TEMLOP-Model and are briefly reviewed in section 2 of this paper. In section 3, we give a full description of the mechanical model and identify the required parameters to describe the cable layout adequately. Then, in section 5, we present the results of calculations with parametric variations in cabling pattern, void fraction and strand stiffness. These results are discussed in section 6, leading to a number of detailed recommendations. The symbols used in the equations are listed in Table I.

Table I. List of symbols.

| | |
|---|---|
| $A$ [m$^2$] | strand cross section area |
| $A_c$ [m$^2$] | projected strand to strand contact area |
| $A_{c\perp}$ [m$^2$] | minimum projected contact area for a strand crossing |
| $B$ [T] | magnetic field |
| $\cos^{-1}\theta$ | cos $\theta$ factor representing the actual cable cross section |
| $D$ [m] | cable width |
| $d_f$ [m] | diameter filamentary region |
| $d_s$ [m] | diameter wire |
| $E_\perp$ or $E_{tr}$ [GPa] | Young’s modulus of the strand, $E$ (transverse) |
| $E_{\|\|}$ [GPa] | Young's modulus of the strand, $E$ (axial) |
| $F_c$ [N] | transverse contact load |
| $f_{cbm}$ [m] | maximum possible average deflection of the cable |
| $f_{cbm}$ [m] | maximum possible deflection available for bending |
| $F_{max}$ [N] | transverse peak load |
| $f_{mc}$ [m] | maximum cable-compression at $F_{max}$ |
| $F_n$ [N] | transverse load with $n$ indicating the order ($n$=1 for $L_{w1}$) |
| $f_{sb}$ [m] | strand deflection from bending, bending amplitude |
| $f_{sb,n}$ [m] | strand deflection from bending, $n$ indicating the order ($n$=1 for $L_{w1}$) |
| $f_{sbm}$ [m] | maximum bending amplitude |
| $f_{sc}$ [m] | strand deformation per strand crossing and line contact |
| $f_{sm}$ [m] | maximum possible average deflection per strand |
| $I_a$ [m$^4$] | moment of inertia |
| $I_c$ [A] | critical current |
| $I_{c0}$ [A] | critical current before loading, virgin state |
| $I_s$ [A] | strand current |
| $J_c$ [A/m$^2$] | critical current density |
| $k_\theta$ | $\cos^{-1}(\theta)$ correction factor |
| $L_w$ [m] | characteristic bending wavelength |
| $L_{w,n}$ [m] | characteristic wavelength with $n$ indicating the order ($n$=1 for $L_{w1}$) |
| $L_\varphi$ [m] | is the characteristic length, $4.3\cdot10^{-3}$ m |
| $N_l$ | number of strand layers |
| $N_s$ | number of strands in the cable |
| $n$-value | value characterizing the steepness of the $V$-$I$ curve [2] |
| $V$ [V] | voltage |
| rho [Ωm] | matrix interfilament resistivity |
| $vf$ | cable void fraction |
| $W_b$ [m$^3$] | section factor written as $d_s^3\cdot\pi/32$ |
| $B_{max}$ [Pa] | maximum magnetic field in the cable |
| $B_{min}$ [Pa] | minimum magnetic field in the cable |
| $\varepsilon$ | axial strain in the $Nb_3Sn$ filaments |
| $\varepsilon_b$ | bending strain in the $Nb_3Sn$ filaments |
| $\varepsilon_{bo}$, e peak-fil reg | peak bending strain in the $Nb_3Sn$ filaments |
| $\varepsilon_{th}$ | thermal pre-compression of the $Nb_3Sn$ filaments |
| $\varphi$ | angle between crossing strands |
| $\theta$ | angle between strand and conductor longitudinal axis |
| $\sigma$ [Pa] | contact stress |
| $\sigma_{max}$ [Pa] | maximum contact stress in the cable |
| $\sigma_{min}$ [Pa] | minimum contact stress in the cable |

## 2. Strand data from TARSIS

In the TARSIS setup, the influence of various loads and deformations (axially tensile, bending strain, contact load from crossing strands or homogeneous transverse load) that frequently occur in a CICC, are studied with different probes [2,12-21]. Over the last few years, there is an increasing interest to study the impact of mainly bending strain on the transport properties of superconducting wires [7,22,23]. The crucial advantage of the TARSIS setup is its ability to measure both the amplitude of deflection or deformation and the applied force with high precision, enabling a full axial and transverse stress-strain analysis.

Lately we presented the results of the probe for periodical strand bending with cyclic loading using different bending wavelengths [21]. In addition, we completed probes for the characterisation of axial tensile stress-strain behaviour of strands [18] and sub-size cables [19]. Since it is the stress distribution, - originating from differential thermal contraction and from the electromagnetic load - that drives the final strain distribution, the stiffness of the strand and cable are definitely key parameters in this study.

Recently, we completed a probe to study also the influence of transverse loading on perpendicularly crossing and pinching strands [17]. While bending is broadly studied, the impact of local transverse loads at strand crossings has hardly received any attention, even though these stresses can be quite severe in the part of the cable cross-section where the load accumulates.
The design of the TARSIS setup is presented in [12] while various experimental results obtained on powder in tube-, internal tin- and bronze-processed $Nb_3Sn$ wires are reported in [13-21].

## 3. Mechanical model for transverse load

### 3.1. General model assumptions

The TEMLOP-Model describes mechanical strand interactions within the cable, including strand bending, strand crossing and line contacts under the influence of an electromagnetic load. The plastic deformation of strands is not separately distinguished although the overall quantitative analysis is improved by accounting for yielding in both, transverse and axial direction, at higher stress levels. This is incorporated in the description for the axial Young's modulus [18]. Thus, a way is found to optimise the quantitative accuracy for the outcome of the model. We do not take into account axial strain variations, other then already included in the bending strain model, so a possible influence of temperature variations and conduit material choice with different thermal expansion is not studied here. The main input for the equations in the TEMLOP-Model is listed in Table II. The input parameters are based on the design layout of the CS1 type of conductor from the CSMC [8], while the strand properties are used from the TFMC strand manufactured by EM (Italy), presently OKCI [20,24].
In the model, the shape of the cable cross section is simply taken square instead of round and no central channel is present. Although it is not complicated to account for the circular cross-section of the actual cable in the numerical model, preliminary estimates showed that the influence on the outcome is not significant. Omitting the central channel from the model corresponds to considering it as a fully stiff medium that transfers the reaction force and displacement without any intrusion.

Table II. Input parameters for the TEMLOP-Model.

| Input parameters strand | | |
|---|---|---|
| Diameter wire, $d_s$ | 810 | μm |
| Diameter filamentary region, $d_f$ | 660 | μm |
| Bending wavelength, $L_w$ | 6 | mm |
| E modulus strand (axial), $E_{\|\|}$ | 29 | GPa |
| E modulus strand (transverse), $E_\perp$ or $E_{tr}$ | 3.3 | GPa |
| **Input parameters cable** | | |
| Strands, $N_s$ | 1152 | strands |
| Layers, $N_l$ | 34 | layers |
| cable void fraction, $vf$ | 0.36 | |
| Cos $\theta$ factor ($\cos^{-1}\theta$) | 1.05 | |
| Cabling scheme CS1 Model Coil conductor | 3x4x4x4x6 | |
| Cabling pitches CS1 Model Coil conductor | 45x74x123x160x380 | mm |
| Cable-compression @ $F_{max}$ measured in Cable Press, $f_{mc}$ | 1.134 | mm |
| magnetic field, $B$ | 12 | T |

### 3.2. Principle of the mechanical model for load distribution

The local transverse deformation of the strands in a CICC is composed of a distribution of bending, contact and homogeneously distributed loads. The local spatial periodic bending, causing a wave-pattern [2,12,13] can actually be described by a periodic repetition of a point load with clamped ends, schematically shown in Figure 1.

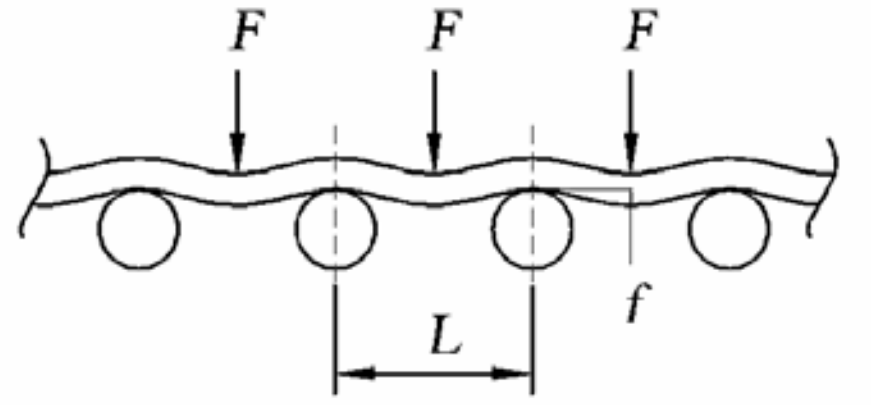

*Figure 1. Schematic view of a point load with clamped ends.*

The corresponding bending beam equations are:

$$f_{sb} = \frac{F \cdot L_w^3}{192 \cdot E_{//} \cdot I_a} \qquad [\mathrm{m}] \qquad (1)$$

$$\varepsilon_b = \frac{F \cdot L_w}{8 \cdot E_{//} \cdot W_b} \qquad [-] \qquad (2)$$

in which $f_{sb}$ is the strand deflection, $L_w$ is the bending wavelength (Figure 1), $E_{//}$ is the Young's modulus in axial direction, $I_a$ is the moment of inertia while $W_b$ is the section factor written as $d_s^3 \cdot \pi/32$ with $d_s$ is the strand diameter. The load $F$ represents the local accumulated magnetic load on a strand in a given layer of the CICC.
The basic model for the strand mechanical interaction between the strands is schematically represented in Figure 2 where three layers of strands (A,B,C) are depicted to illustrate the deformation, for the load in three steps for increasing load from Figure 2a to c.

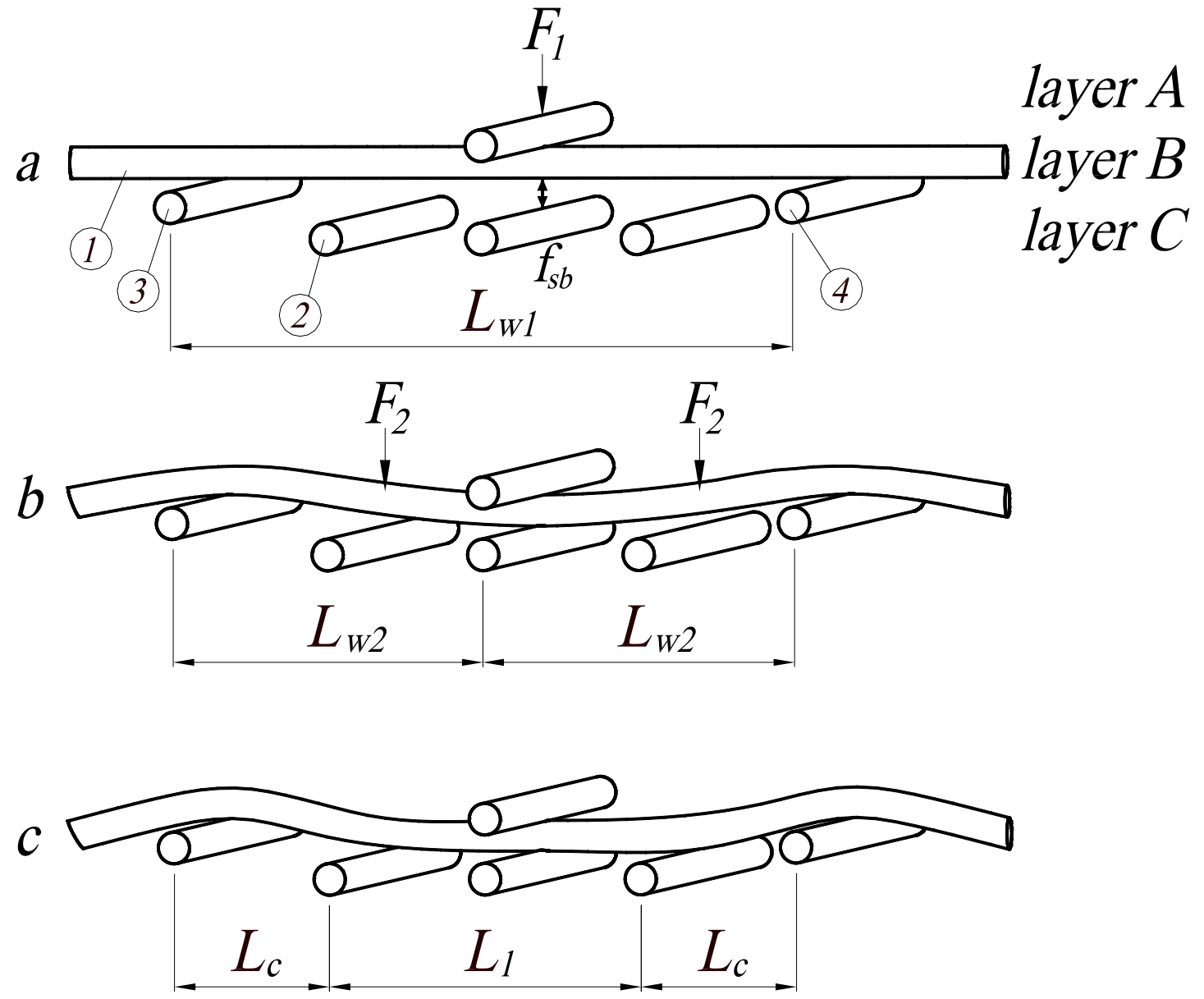

*Figure 2. Schematic model for strand mechanical interaction in the cable bundle from layer to layer with increasing electromagnetic force pointing from top to bottom.*

The first case (Figure 2a) corresponds to the virgin situation before electromagnetic loading of the cable bundle. The second scheme in Figure 2b shows the situation when the deflection of strand 1 in layer B, characterised by the bending amplitude $f_{sb1}$, has virtually reached its maximum value and strand 1 is about to touch the surface of strand 2 in the layer C below. This situation constitutes a limiting case for bending with the wavelength characterised by $L_{w1}$. The strand (1) with the wavy pattern is still only supported by the crossing strands associated with the wavelength indicated with $L_{w1}$, i.e. the transverse contact load on the strand crossings remains concentrated on the same number of strand crossings that were already present from zero load, on a distance $L_{w1}$. In this situation, the bending is fully described by the point load $F_1$, the wavelength $L_{w1}$ and the maximum bending amplitude ($f_{sbm1}$). The transverse stress is transmitted through the contact points between strand 1 and the two supporting strands 3 and 4.
With further increasing the load (Figure 2b), strand 1 with the wavy pattern will make contact with strand 2. Arriving at this stage the periodicity in loading contacts increases by a factor two, while the maximum possible deflection ($f_{sb2}$) becomes half the initial value and the load $F_2$ becomes half $F_1$. Hence, we can distinguish for a decrease by a factor two of the characteristic bending wavelength from $L_{w1}$ to $L_{w2}$ and a doubling of the number of contact loads. The transverse load, initially a reaction force only supported by the contacts spaced by the original periodicity in crossing strands represented by $L_{w1}$, becomes also partly distributed now by strand 2 from layer C below. The decrease in wavelength from $L_{w1}$ to $L_{w2}$ has two important effects: it restricts further deflection (see relations (1) and (2) and it causes a more homogeneous distribution of the contact stress and point loads.

When the load is further increased by other strands in layer A, (not indicated in the schedule) strand 1 will make contact with other strands from the same layer (C) as to which strand 2 belongs (Figure 2c). The result is a further successive decrease of the bending wavelength introducing now the more general description $L_{w,n}$, the point load $F_n$ and the maximum possible deflection $f_{sb,n}$. At the same time a progressive increase of the contact area occurs, which furthermore becomes more homogeneously distributed. Both factors reduce the local peak contact stress.
Now that the basic model for strand mechanical interaction has been introduced, we can derive the maximum bending amplitude and strand deformation at the crossovers.
The peak strain occurs at the strand crossings with the point loads. With reaching the maximum deflection and going from $L_{w,n}$ to $L_{w,n+1}$, the peak strain at the initial point load $F_1$ will not increase further (may actually decrease) but new maxima are created at other locations corresponding with the position of the $F_2$ point loads. To obtain the strain value that limits the critical current in a strand, we take the peak strain along the strand.

### 3.3. Strand and cable deformation

The accumulation of Lorentz load $F_n$ on the strands in a CICC is further explained in Figure 3 where the electromagnetic *IxB* load increases progressively from the right to the left. At the same time, the local magnetic field decreases also in this direction, due to the combination of the applied field generated in the coil (or background field) and the cable self-field due to the transport current.

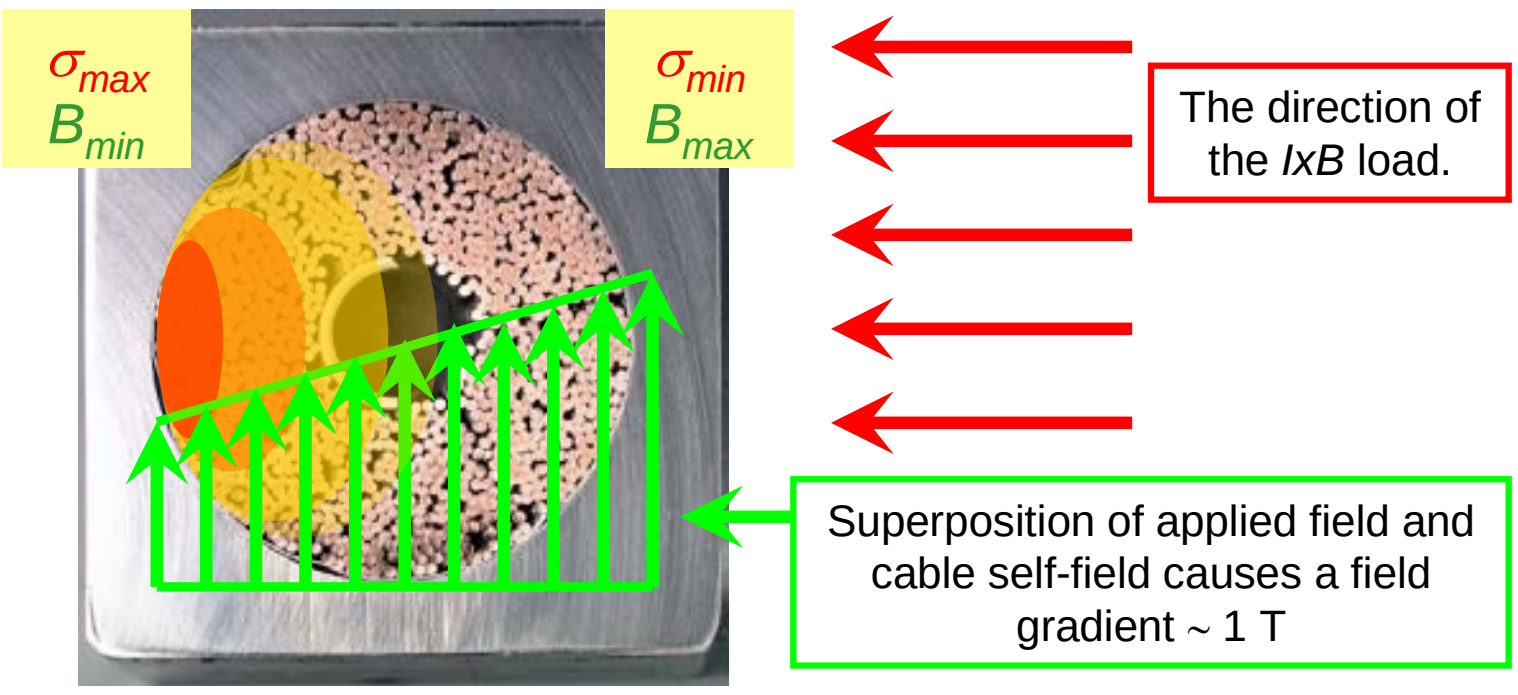

*Figure 3. The magnetic field profile in the cable cross section in relation to the accumulated load distribution.*

The locally occurring accumulated peak load in an ITER conductor with 1152 strands is about 20 kN/m, assuming that the number of strand layers can be approximated by the square root of the total number of strands ($N_s$) in a cable [2,6]:

$$F = I_s \cdot B \cdot N_s^{0.5} \qquad \text{[N/m]}, \qquad (3)$$

As already indicated in [25], the most favourable transverse load distribution is obviously a homogeneous one, corresponding to a situation with infinite line contacts between parallel strands and avoiding local bending effects. In that case, the transverse peak stress, $\sigma_{max}$, is the locally occurring accumulated peak load in an ITER conductor of 20 kN/m on the projected area of a strand cross section ($d_s$).

$$\sigma_{hom} = \frac{I_s \cdot B \cdot N_s^{0.5}}{d_s} \qquad \text{[Pa]}, \qquad (4)$$

The peak transverse stress in this homogeneous case, $\sigma_{hom}$, amounts to ~20 MPa. However, as soon as there is locally sufficient space available for strand bending, the load at the crossovers rises and will exceed the value determined for a homogeneous distribution, since the projected contact area $A_c$ becomes smaller than in the case of an infinite line contact. The projected contact area between two crossing strands depends on the angle between the wires, $\varphi$, and on the strand diameter:

$$A_c = \frac{d_s^2}{\sin\varphi} \qquad [\text{m}^2] \qquad (5)$$

Correspondingly, the local transverse stress is multiplied with a contact area geometrical factor $k$:

Superc. Science & Techn. 21 April 2006.

$$k = \frac{d_s \cdot L_w}{A_c} = \frac{L_w \cdot \sin\varphi \cdot}{d_s} \qquad [-] \qquad (6)$$

Combining (4), (5) and (6) we can write for the local peak stress:

$$\sigma_{\max} = \frac{\sigma_{\text{hom}} \cdot L_w \cdot \sin\varphi}{d_s} \qquad [\text{Pa}] \qquad (7)$$

which can also be written as:

$$\sigma_{\max} = \frac{I_s \cdot B \cdot N_s^{0.5} \cdot L_w \cdot \sin\varphi}{d_s^2} \qquad [\text{Pa}] \qquad (8)$$

Apart from the electromagnetic load and strand mechanical properties, the associated bending deflection is also determined by the distance between the supporting strands (relation 1). This wavelength is in turn determined by the cabling pattern of the strand bundle and is connected to the number of elements per sub-cable and the twist pitches and finally by the load determining $L_{w(n)}$.

The actual average wavelength ($L_w$) for crossing strands (and bending) is determined from a disassembled full-size ITER CS1 conductor and is 6 mm with a standard deviation of 2 mm, while the crossing angle amounts to 45 degrees. For a so-called SeCRETS-A sub-size conductor with a practically identical cabling scheme, also a wavelength of 6 mm was found [26]. A photograph of the CS1 cable bundle in Figure 4 shows clearly the deformation at strand crossings and varying periodicity.
A conductor with longer pitches, a CS2 type of conductor used for the outer module of the CSMC, is shown in Figure 5. For the CS2 type of conductor, the average $L_w$ is difficult to determine as the positions of the crossings are less clear but it is defivitely longer than 6 mm while the crossing angle $\varphi$ amounts to about 10 degrees.

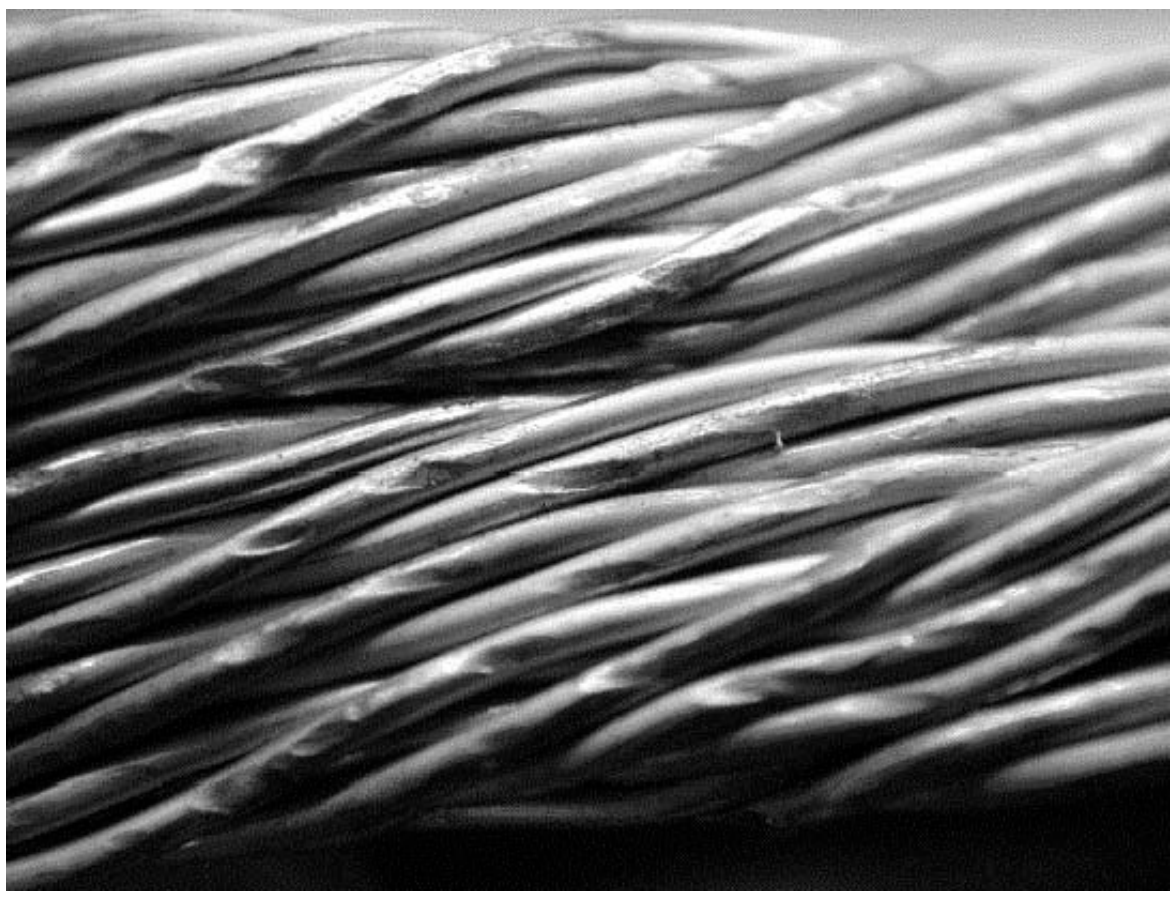

*Figure 4. The cable structure of an ITER CS1.1 Central Solenoid Model Coil conductor.*

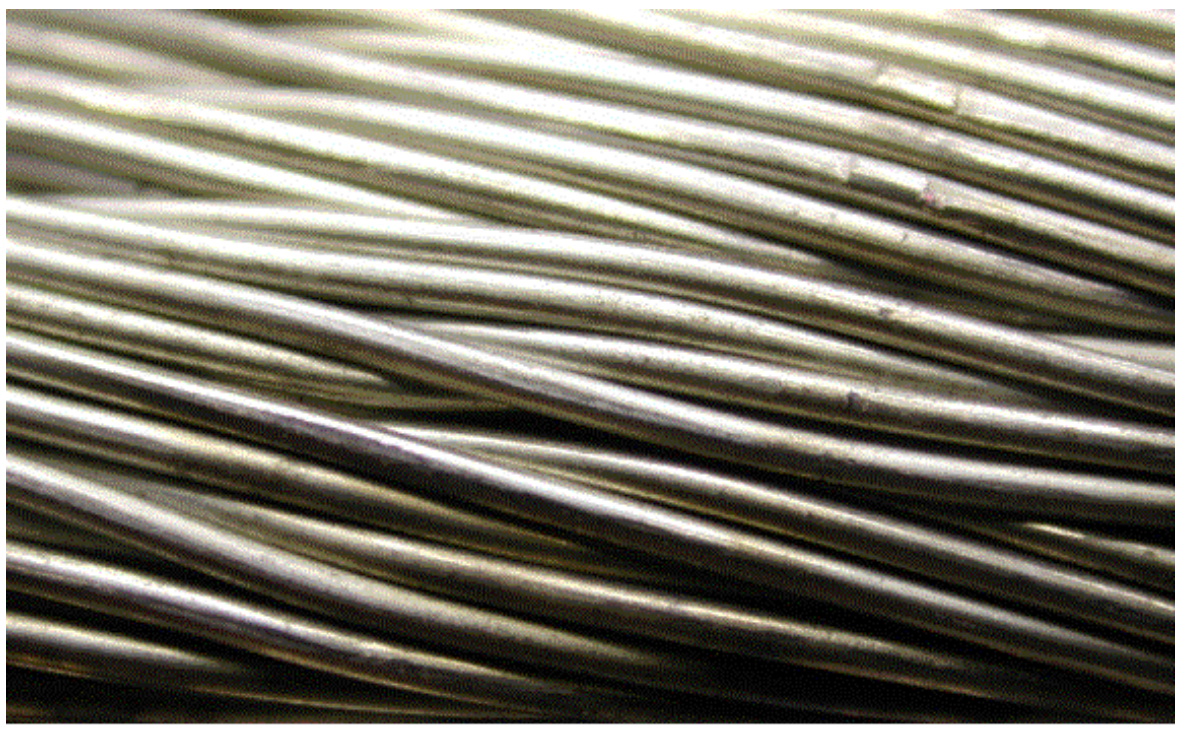

*Figure 5. The cable structure of an ITER CS2 Central Solenoid Model Coil conductor with longer twist pitches.*

This implies that for a wavelength of 6 mm the accumulated load per strand crossing amounts to about 100 N. For a periodicity of $L_w$=6 mm of the CS1 conductor and a strand diameter of 0.81 mm, the local peak stress can then reach a maximum level of 110 MPa.

Due to the cabling scheme in the ITER CICCs (Table II) in several cabling stages with subsequent twist pitches it seems unavoidable to have local stress/strain concentrations at the crossover contacts and bending in between them. The bending deflection itself is restricted in absolute sense by the average free space available in the cable bundle (Figure 6). When the cable bundle is represented by a square bundle of strands with $N_l$ layers of strands and each layer containing $N_l$ strands (in this case $N_l$=34) we can derive the maximum possible free distance for a strand to move before encountering a neighbouring strand. The numbering of the strand layers $N_y$ goes from 1 to $N_l$, starting at the high field (low *IxB*) side of the conductor.

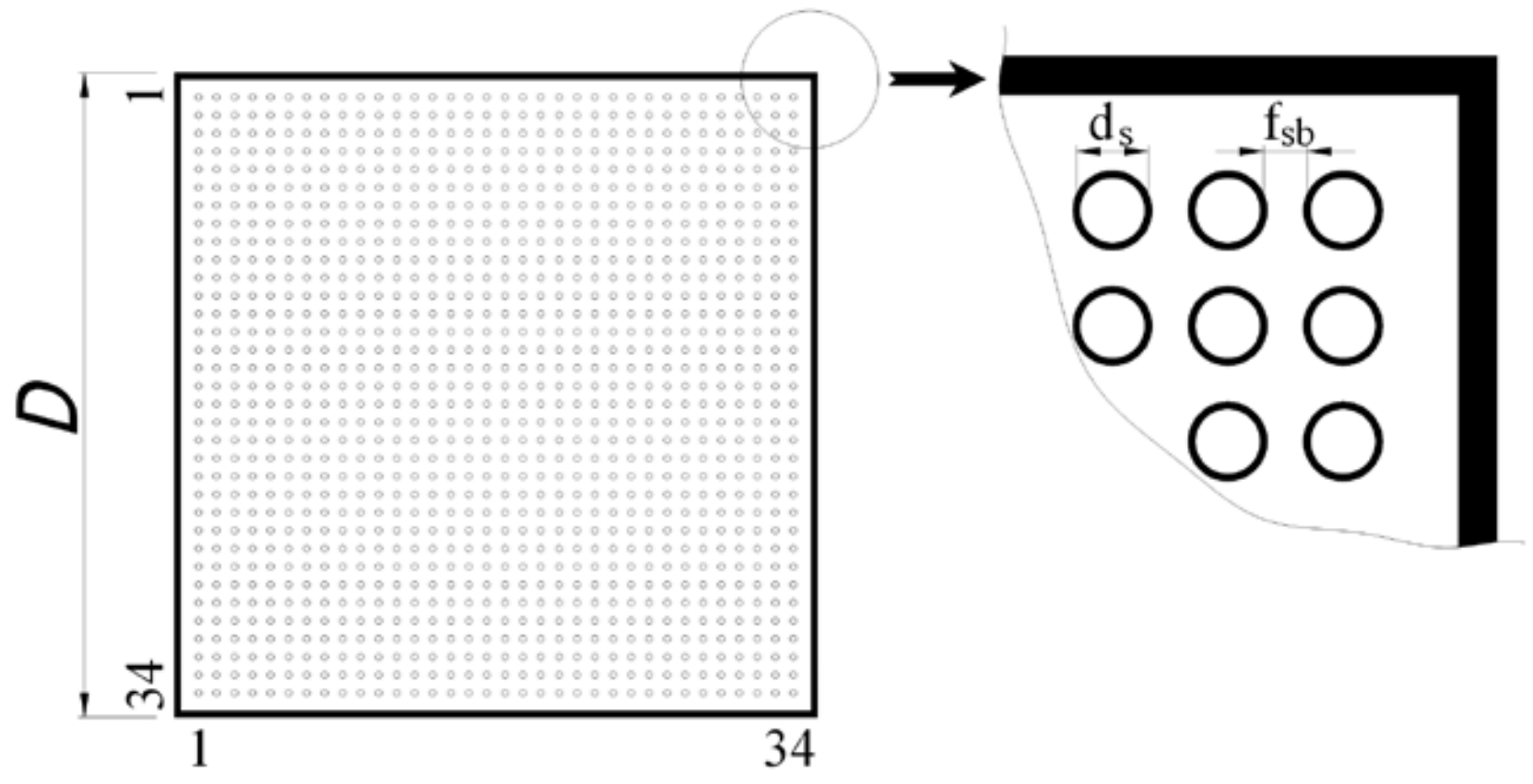

*Figure 6. Schematic representation of the mechanical model for cable compression to derive the initial free deflection for bending per strand.*

For a given void fraction *vf*, number of strands $N_s$ and strand diameter $d_s$, the cable width *D* is calculated with:

$$D = \left( \frac{N_s \cdot \pi \cdot d_s^2}{4 \cdot (1 - vf) \cdot \cos\theta} \right)^{0.5} \qquad [\text{m}] \qquad (9)$$

The maximum possible deflection available for bending for $N_l=(N_s)^{0.5}=34$ ($f_{cbm}$) is derived from the cable width *D,* the strand diameter $d_s$ and the cos*(θ)* factor, correcting the strand cross-section area for the average angle between the twist pitches in the cabling with respect to the cable axis:

$$f_{cbm} = D - N_s^{0.5} \cdot d_s \cdot \cos^{-0.5}(\theta) \qquad [\text{m}] \qquad (10)$$

The average cos*(θ)* factor from the CS1 type of conductor in Figure 4, with angles for $\theta$ between roughly 15 and 20 degrees, is determined on 0.95 [27]. The average angle $\theta$ decreases with an increasing cabling pitch length, and is correlated to the cable void fraction. We assume that the cos*(θ)* correction factor rises by approximation linearly with the increase of the cable twist pitches and at the same time that the characteristic bending wavelength $L_w$ also increases linearly with the twist pitch. As we want to use the $L_w$ as a variable input parameter for the model we define the $\cos^{-1}(\theta)$ correction factor $k_\theta$ as:

$$\cos^{-1}(\theta) = k_\theta = 1 + \frac{\lambda}{L_w} \qquad [\text{-}] \qquad (11)$$

with $\lambda$=3·10-3 m, empirically derived from the CS1 type of conductor, for $k_\theta$ = 1.05 and $L_w$=0.006 m.

When we fix the void fraction as a conductor parameter, the space between the strands for maximum possible deflection (for bending) will change with varying wavelength, following relation 11.

The maximum possible average deflection per strand ($f_{sm}$) is then calculated from the entire cable ($f_{cbm}$) as:

$$f_{sbm} = \frac{f_{cbm}}{N_s^{0.5}} \qquad [\text{m}] \qquad (12)$$

For the calculation of the maximum free deflection per strand we take the two dimensional free space direction. If we allow for a 2D free space displacement, multiplication of the one-dimensional $f_{sbm}$ by a factor $(2)^{0.5}$ is required. The real maximum possible deflection of the strands however, is expected less than for a regular grid of parallel strands with cos*(θ)=1*, as schematically presented in Figure 6. In a real cable, the spatial distribution of strands is not that regular at all. Most of the free space in a cable, determining the void fraction, seems concentrated at larger gaps likely between the higher cabling stages (not the final-one) and is not available for the majority of the individual strands. This is illustrated by the CS1 cable cross section in Figure 7, where larger voids are surrounded by regions with rather highly compacted strands. Although this conductor was mechanically loaded in the press, virgin conductors show the same pattern of larger void concentrations.

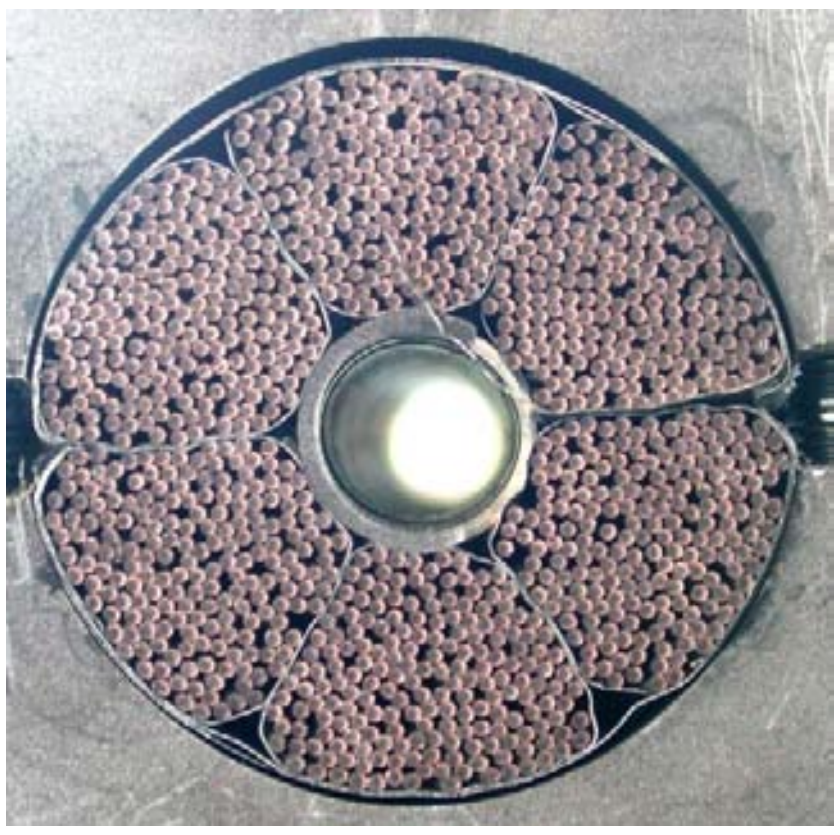

*Figure 7. The cable cross section of an ITER CS1.1 Central Solenoid Model Coil conductor after transverse cyclic loading in the Twente Cable Press.*

Together with the variation in the wavelengths, the angle of the strands with the surrounding strands and the friction between the strands (causing hysteresis) the theoretically maximum deflection becomes further restricted. Based on the photographs of conductor cross sections we estimate the average available space between strands and we correct the theoretically maximum deflection per strand ($f_{\mathrm{sbm}}$) by a factor of 3 and determine the effective $f_{\mathrm{sbm}}$ by:

$$f_{sbm} = \frac{f_{cbm}}{3 \cdot N_s^{0.5}} \text{ [m]} \tag{13}$$

For the CS1 type of cable with a void fraction of 0.36 and 6 mm wavelength, we obtain a maximum (average) possible bending deflection per strand ($f_{\mathrm{sbm}}$) of 41 μm. This is still without accounting for contact deformation at the strand crossings but this mechanism is built-in further on.

Transverse stress-strain measurements are performed with the TARSIS crossing-strands probe on the LMI-EM strand [20] and are presented in Figure 8. It appears that the loading curve is practically without hysteresis when unloading. The initial deformation requires only a very small load and we assume that the initial strand deformation up to a diameter compression of almost 40 μm (5 % strain) already occurs partly during cabling of the conductor, but mostly throughout the final compacting step, after inserting the cable in the conduit. This allows us to use a constant value of 3.3 GPa for the strand transverse modulus of elasticity, $E_{\perp}$ in the model.

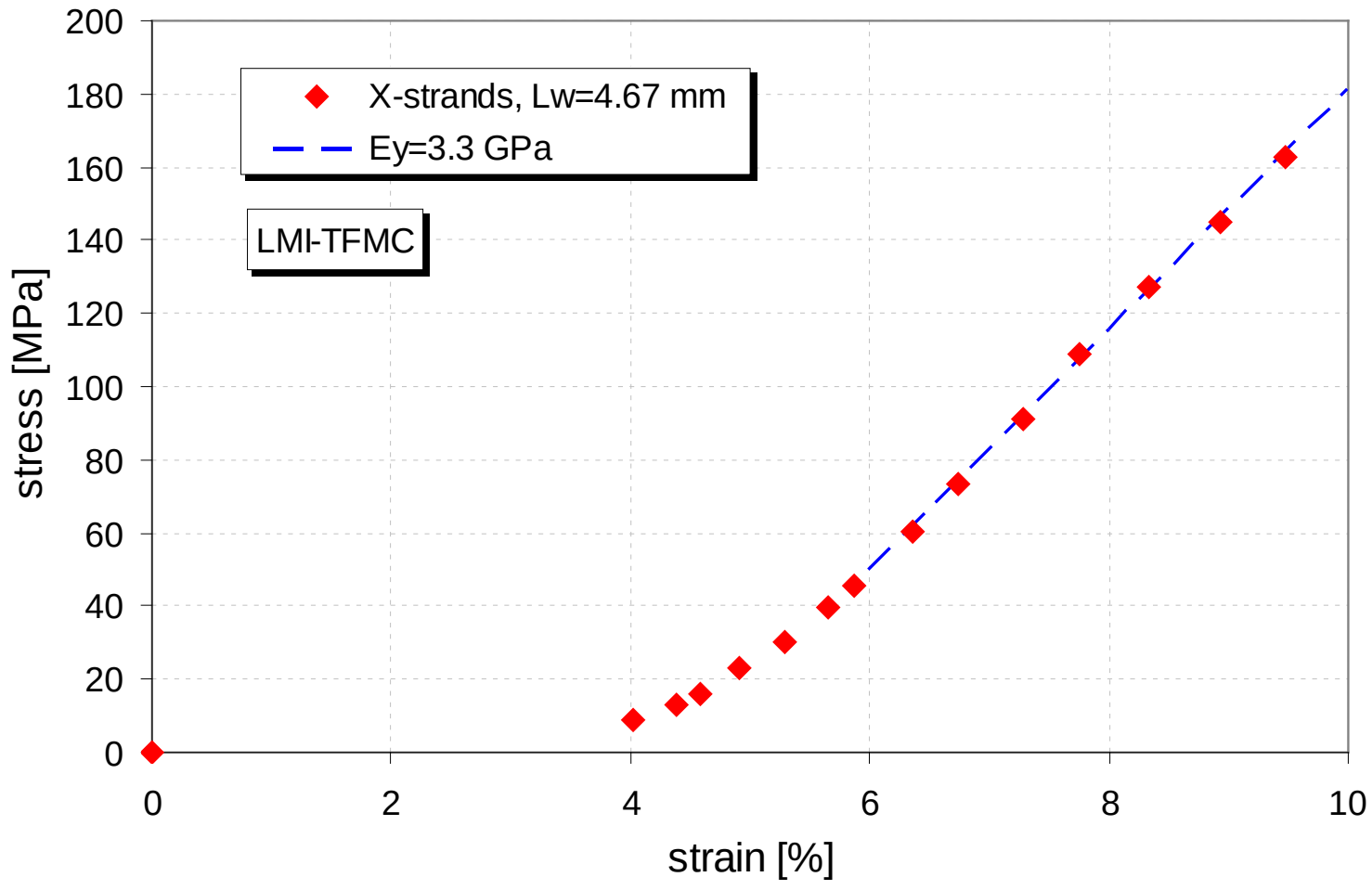

*Figure 8. The stress-strain characteristic of the LMI-EM strand measured in the TARSIS crossing strands probe [20].*

The actual available space for bending in the virgin state is determined by the calculated $f_{sbm}$ (relation 13). However, when the cable is compressed by the electromagnetic load (or mechanically in the press), the strands will deform not only by bending but also at the crossings and line contacts. As a result, the available space for bending will be further restricted. This is formulated in the model as:

$$f_{sb} = f_{sbm} - f_{sc} \qquad \text{[m]} \qquad (14)$$

with $f_{sc}$ the strand deformation per strand crossing and line contact and $f_{sb}$ is the space for bending, decreasing with accumulating electromagnetic load from layer to layer in the cable bundle.
The bending mechanism throughout the layers of the cable, is now described with relations 1, 2 and 9 and the successive limitation of the possible deflection $f_{sb\text{-}n}$ with increasing $n$ for higher orders of $L_w$, limiting the possible peak strain in the strands with relation 14.

The basic principle of contact stress evolution in the TEMLOP-Model is depicted in the scheme of Figure 2, where strand 1 from layer B is crossing under an angle $\varphi$ with the strands belonging to layers A and C. With further increasing the load (Figure 2b), strand 1 with the wavy pattern will make contact with strand 2. The transverse load, initially a reaction force only supported by the contacts spaced by the original periodicity in crossing strands represented by $L_{w1}$, becomes also partly distributed now by strand 2 from layer C below. Important is also the angle $\varphi$ in determining the contact stress and this angle is directly connected to the cabling pitches (Figure 4 and Figure 5). We assume that the relation between $L_w$ and the crossing angle $\varphi$ is linear and inverse proportional for larger wavelengths. Based on the crossing angle $\varphi$ and $L_w$ determined for the mentioned CS1 and CS2 type of conductors we obtain empirically:

$$L_w \cdot \sin(\varphi) = 4.3 \cdot 10^{-3} \qquad \text{[m]} \qquad (15)$$

in which $4.3 \cdot 10^{-3}$ m is the characteristic length $L_\varphi$. So, instead of relation 5, we can write now for the contact area, $A_c$:

$$A_c = \frac{A_{c\perp} \cdot L_w}{L_\varphi} \qquad \text{[m}^2\text{]} \qquad (16)$$

with $A_{c\perp}$ as the minimum projected contact area for a strand crossing. Towards shorter wavelengths, the absolute minimum of this contact area is restricted by the square of the strand diameter, $d_s$. Thus, we use a hyperbolic description for the computation of the effective contact area in relation to $L_w$:

$$A_c = d_s^2 \left( \frac{L_w^2}{L_\varphi^2} + 1 \right)^{0.5} \qquad \text{[m}^2\text{]} \qquad (17)$$

This description of the contact area is used for the computation of the results presented in this paper. The numerical method uses discrete steps in the increase of the contact area from $L_{w(n)}$ to $L_{w(n+1)}$, which is the explanation for the discontinuities in some of the curves shown later on. The $A_c$ increases stepwise from $L_{w(n)}$ to $L_{w(n+1)}$, proportional to $n$ instead of doubling each time the contact area in order to make the strand layer interactions more gradual. This can be considered as a conservative approach with respect to the influence of contact stress.

Instead of using the discrete steps from $L_{w(n)}$ to $L_{w(n+1)}$ in the numerical process for the increase of the contact area with rising load as explained above, we can also define a more gradual evolution of this process. In the basic principle of the TEMLOP-Model depicted in the scheme of Figure 2, strand 1 from layer B is crossing under a relatively large angle $\varphi$ with the strands belonging to layers A and C. It is not difficult to imagine that for a much smaller crossing angle, the crossing strands from layer C act nearly as a line contact for strand 1 from layer B. This is well illustrated by the cabling pattern in Figure 5.

The effect of a limited bending deflection, further restricted by the deformation at the strand contacts at increasing load, combined with a progressively homogenisation of the contact load can be formulated as follows:

$$A_c = \left(L_{w_n} - 2 \cdot L_c\right) \cdot \frac{d_s}{\sin\varphi} \qquad [\mathrm{m}^2] \qquad (18)$$

in which $L_c$ is the (decreasing) "characteristic bending length" as indicated in Figure 2c and $L_l$ is the (increasing) length along strand 1 with (homogenously) distributed load: $L_l$=$L_w$-2.$L_c$. The minimum value of $A_c$ is given by relation 5 (with $\varphi$=45° for $L_w$=6 mm) starting in the virgin condition of the cable, represented by Figure 2a. As soon as strand 1 meets strand 2, not only the bending wavelength becomes shorter but also at the same time, $A_c$ becomes larger. The characteristic bending length $L_c$ is derived from the standard bending beam formula for a point load with one clamped end and a supported end at the other side but obviously also other descriptions can be used.

$$L_c = \left(\frac{48\sqrt{5} \cdot f_{sb} \cdot E_{//} \cdot I}{F}\right)^{1/3}, \quad [\mathrm{m}] \qquad (19)$$

It appeared that utilising relation 19 leads to similar conclusions as when using relation 17.

With mentioned relations 1-17 and the relevant strand parameters like the experimentally determined axial and transversal $E$-modulus (and diameter), the overall mechanical response of the cable to an electromagnetic force can be described now. For the cable properties, we need the void fraction to determine the bending deflection limit. The $L_w$ is a parameter that is directly corresponding to the cabling pitches and will be varied to demonstrate the influence on the conductor performance. The conductor performance in terms of critical current reduction due to bending and contact stress is included by means of experimentally determined relations that are presented hereafter.

## 4. Critical current reduction in strands

### 4.1. Bending strain

It was recognised by Ekin [1], that the transport properties of an $Nb_3Sn$ strand is not only affected by the applied bending strain but also depends on the inter-filament electrical resistivity. The electrical resistance between the twisted filaments determines whether the distance between the periodically distributed peak strains in the filaments is short or long compared to the current transfer length. One extreme is that current transfer between the filaments is not allowed due to the high resistance, and then the minimum $I_c$ for each filament specifies the filament critical current. In this case, the strand $I_c$ corresponds to the sum of the filament critical currents, which is limited by the maximum strain along filaments at any point. In this extreme case for short pitch or high matrix resistivity (rho), each $I_c$ is written as [1]:

$$I_c = \frac{2A}{\varepsilon_{b0}^2} \int_0^{\varepsilon_{b0}} J_c(B, T, \varepsilon_{th} - \varepsilon)\varepsilon d\varepsilon \qquad [\mathrm{A}] \qquad (20)$$

$A$ is the strand cross section area, $\varepsilon$ is the strain in the filaments over a cross section of the strand, $\varepsilon_{bo}$ is the peak bending strain in the outer ring of filaments, $\varepsilon_{th}$ is the thermal pre-compression of the $Nb_3Sn$ filaments and $J_c$ is the critical current density. Note that for a strand the position of the filaments is fixed on a certain radius when travelling along the axis of the wire while in a CICC the strands are fully transposed.

If on the other hand current transfer is allowed at a low voltage level, the overall $I_c$ of a strand is the sum of the filament currents at any section considering the local strain variation over the section. In this other limiting case for long pitch or low matrix resistivity, $I_c$ can be expressed as:

$$I_c = \frac{2A}{\pi \varepsilon_{b0}^2} \int_{-\varepsilon_{b0}}^{\varepsilon_{b0}} J_c(B,T,\varepsilon_{th}+\varepsilon)\sqrt{\varepsilon_{th}^2-\varepsilon^2}\,d\varepsilon, \qquad [A] \qquad (21)$$

In order to solve the equations 20 and 21 one needs to know the axial $I_c(\varepsilon)$ relation, which can be obtained experimentally [28]. The shape of the computed curves from relation 20 and 21 is primarily driven by the $I_c$-strain variation of the particular $Nb_3Sn$ strand and are both plotted in Figure 9. It was clearly identified recently, at least for several of the ITER Model Coil strands used in the CSMC and TFMC, that the reduced critical current can be accurately described by only the low resistivity regime representing full interfilament current transfer. [21] (see Figure 9). The data are measured with an $I_c$ criterion of 10 µV/m. However, experiments were performed on different strands that were swaged into a steel tube [14], to simulate the influence of the axial thermal pre-compression of the conduit on the cable bundle due to the higher coefficient of thermal expansion. It appeared for these strands that the reduced $I_c$ does not correspond anymore to the low transverse resistivity regime [16] although the internal resistivity remains the same. An example is given in Figure 10, where a bronze strand indicated with VAC (from CS1-type) which was used for the inner layer CSMC production, showed such a typical performance. The reduced $I_c$ versus bending strain is plainly lower then predicted by relation 21 and it is not clearly understood why, although strand FEM simulations performed by Mitchell [29,30] suggest an influence of the radial strain component. We confirmed by means of AC loss measurements, that the steel compression did not affect the interfilament resistivity, possibly causing a shift downwards the high resistivity regime [16].

For the TEMLOP-Model, we assume that the real $I_c$ reduction with increasing bending strain under axial pre-compression applies to the uncompressed strand condition (only strand cool-down strain). This is less severe than the performance of the strand in a steel tube (with imposed axial and radial thermal compression) but conservative again with respect to the case from relation 21, representing axial compressed strand. Therefore, we take the polynomial fit through the experimental data in Figure 9 to account for the bending strain in the TEMLOP-Model. In fact, this assumption and the factor 3 from relation (13) mainly determine the quantitative accuracy of the model, but the qualitative prediction in performance remains unaffected.
For the layers of strands travelling from lowest load to the accumulated peak load, the magnetic field decreases from 12 T to 11 T (see Figure 3). As the critical current varies substantially through the cable cross section due to this field variation, we account for this effect by calculating the dependency of the critical current at 11 T with relation 21 and the strand $I_c(\varepsilon)$ data. The local critical current reduction is then determined by the local field inside the cable by a weighted interpolation of the reduction between 11 T and 12 T, assuming a linear field profile in the cable cross-section from layer to layer.

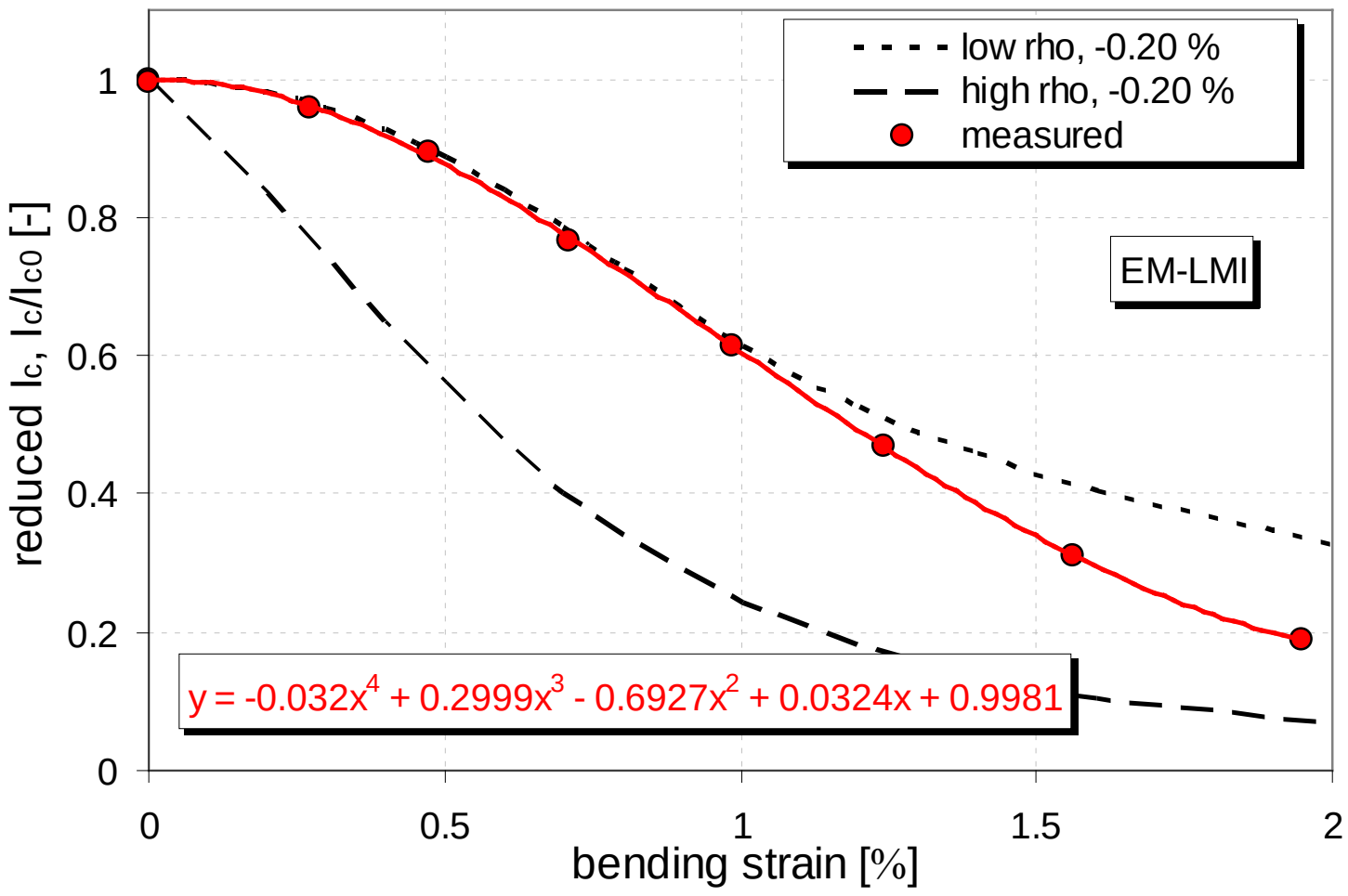

*Figure 9. Measured $I_c$ versus applied bending strain (with polynomial fit) at 12 T and 4.2 K for the LM-EM strand used for the TFMC (taken from [20])with the computation of the curves for full and no inter-filament current transfer under bending.*

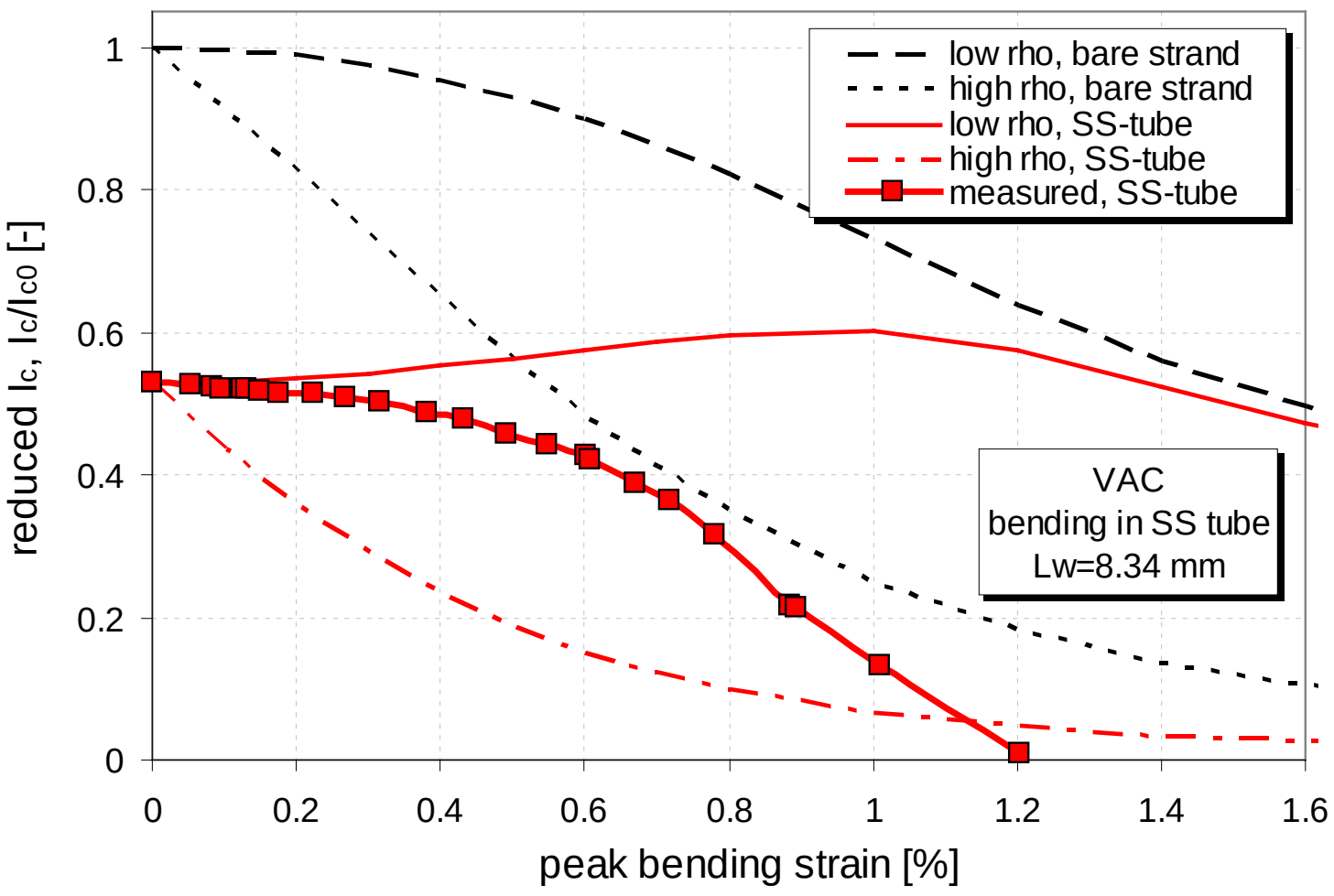

*Figure 10. Measured reduced $I_c$ versus applied bending strain (with polynomial fit) at 12 T and 4.2 K for the VAC bronze strand used for the CSMC (taken from [16])with the computation of the curves for full and no inter-filament current transfer under bending (bare strand and with SS tube).*

### 4.2. Contact stress

A similar strategy is followed for the interaction of the transverse load on strand crossings. Although several experiments are known [31-34], there is no analytical expression available that describes the reduction of the critical current under increasing contact load. On top of that, it is obvious that a polynomial fit through the experimentally obtained dependency has the best accuracy. Therefore, we measured the reduction of the critical current for the LMI-EM strand with the TARSIS probe for crossing strands [20] and the results are depicted in Figure 11.

In addition, also here we account for the magnetic field decrease along the cable cross section from 12 T to 11 T (see Figure 3) and measured the dependency at both fields. Similar to the method applied for strand bending, the local critical current reduction is determined by the local field through weighted interpolation between the measured reduction at 11 T and 12 T through the cable from layer to layer. Note that for the strands that are in 11 T magnetic field in the accumulated load region, the critical current is not affected up to 75 MPa stress.

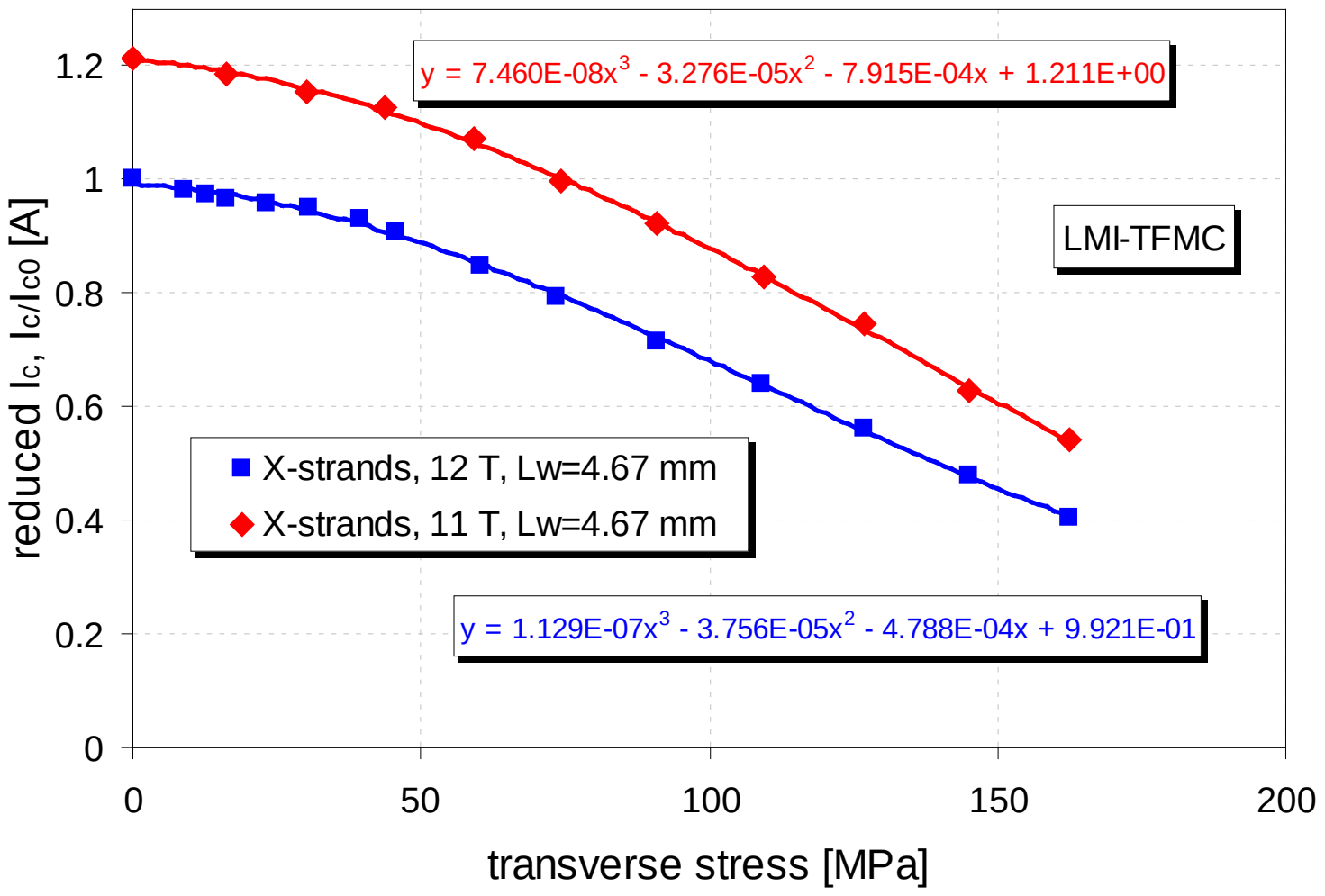

*Figure 11. The reduction of the critical current versus the applied stress for the LMI-EM strand at 11 T and 12 T measured in the TARSIS crossing strands probe [17].*

### 4.3. Integrated model

The overall critical current reduction in the cable is calculated layer by layer from minimum to maximum accumulated load and in a magnetic field changing from 12 T to 11 T. Just as for interfilament current transfer in a strand described by relations 20 and 21, we can distinguish between no and full current transfer between strands in the cable. When we assume full current transfer between strands then the current is allowed to redistribute and we can add all critical currents from layer to layer to get the average cable critical current reduction. This is reflected by the following relation:

$$\left(\frac{I_c}{I_{c0}}\right)_{cable} = \frac{\sum_{1}^{N_l}\left[\sum_{L_{w_1}}^{L_{w_n}}\left(\frac{I_c}{I_{c_0}}(\varepsilon_b, B)\cdot\frac{I_c}{I_{c_0}}(F_c, B)\right)\right]}{N_l} \quad (22)$$

in which the peak bending strain in the filamentary region ($\varepsilon_b$) is derived from the load in the layer of strands $N_l$. The reduced critical current, $I_c/I_{c0}$, depends on the field and if the interpolated value exceeds 1 when the combination of bending strain and magnetic field is sufficiently low not to reduce the real critical current, a value of 1 is just taken instead. The reduced critical current from the transverse contact load ($F_c$), represented by the right term in relation 22, is multiplied with the reduction obtained from the bending strain (left term) assuming that both reductions add at the strand crossovers. The $I_c$ reduction of all layers is averaged to come to the reduced $I_c$ of the entire cable, because we assume full current distribution between strands.

When the current in the strands is not able to redistribute due to high interstrand contact resistance, the peak strain and the corresponding minimum $I_c$ limits the current for all strands as all strands have to pass the peak strain region within one pitch length of the final cabling stage. Depending on the ratio of the consecutive cabling pitches, it may be possible that the peak strain region is passed with a periodicity larger than the last stage cabling pitch. A high interstrand contact resistance, or in other words, a long current transfer length, further limits the $I_c$.

In [35] the influence of the current transfer length is pointed out for a full size NbTi conductor. We found that the current transfer length between the wrapped final-minus-one cabling elements (petals) amounts to several tens of meters. Without the resistive wraps around the last minus one cabling stages the current transfer length amounts to several meters. This means that for strands inside the last-minus-one cabling stage some redistribution may be allowed when the periodicity is larger than one last stage pitch length. A conservative approach would be to choose for no redistribution. For our analysis we examine both limts, full and no current redistribution.

The model can be further refined by changing the stepwise change in the bending wavelengths from $L_{w(n)}$ to $L_{w(n+1)}$ into a more gradual numerical approach but this has likely only minor effect on the overall outcome.

As already mentioned above, the exact influence of the conduit axial thermal pre-compression on the critical reduction of the cable $I_c/I_{c0}(\varepsilon_b,B)$ is not known and only for this part of the model, we rely on a hypothesis with some higher degree of uncertainty. At this point some additional effort is required, asking for a more fundamental approach in investigation of the 3D strain behaviour of $Nb_3Sn$ layers in relation to strand geometries.

## 5 Results of model computations

### 5.1. Mechanical response

The cable deformation (transverse compression) versus the applied electromagnetic *IxB* load is calculated for different wavelengths $L_w$ in a conductor with 1152 strands and a void fraction of 0.36. The strand properties are from the LMI-EM strand with $E_{//}$ = 29 GPa and $E_{\perp}$ = 3.3 GPa [20]. The results are plotted in Figure 12. The cable becomes stiffer with shorter wavelength, which is also illustrated by Figure 13 where the overall transverse cable modulus of elasticity versus the applied electromagnetic *IxB* load is presented for different wavelengths.

The deflection for bending and contact load is depicted in Figure 14, for the total deflection we assume that both components can simply be added although this is not trivial. The TEMLOP-Model gives a total cable compression of about 1.3 mm, with 1.1 mm for bending and 0.2 mm for direct strand contact stress on a cable with a void fraction of 0.36 and $L_w$=0.006 m. The total transverse cable compression saturates for contact and bending stress with increasing *IxB* load. After about 400 kN/m, bending reaches its maximum deflection limit while the contact deformation progressively continues.

The deflection from bending and contact load differs from layer to layer through the cable. In Figure 15 we can see how the deflection starts to increase linearly in the layers from high to low magnetic field until the maximum limit is reached around layer 4, coming in contact with strands in the next layer ($L_w$=0.007 m).

Here the contact stress changes rapidly due to the numerical structure of the model, imposing a sudden increase of the contact surface when changing from $L_{w1}$ to $L_{w2}$. Behind this layer the bending deflection becomes further restricted because the increasing contact stress is compressing further the strand crossovers, at the same time increasing the contact deformation (deforming the strands). At layer 20, the contact stress changes rapidly again due to the sudden increase of the contact surface when changing from $L_{w2}$ to $L_{w3}$.

In Figure 16 we find the typical initial increase of the bending strain from the section characterised by $L_{w1}$=0.007 mm together with the rise of the contact stress until reaching contact with the first strand of the layer below. Then the bending strain linked to $L_{w1}$ (almost 0.7 %) relaxes and the contact stress increases with only half the initial slope. However, at the same layer the bending strain connected with $L_{w2}$ starts to increase, reaching a peak strain of about 1.1 %, exceeding that of the maximum strain connected to $L_{w1}$, but deeper in the conductor and in significantly lower field. This maximum for $L_{w2}$ bending occurs at layer 20, where the maximum deflection is reached for this wavelength. The peak strain related to $L_{w3}$ remains restricted to below 0.5 %. The contact stress reaches its maximum accumulated load at the low field side of the conductor at 40 MPa. We also find that $L_{w4}$ has not yet become involved in the loading process.

Figure 17 shows what happens with the peak strain and stress when we double the $L_{w1}$ to 0.014 mm. We find that $L_{w4}$ has also become involved here but the peak strain occurs now in layer 11 instead of layer 23 and is associated to $L_{w3}$. However, the peak strain in the cable is now less than 0.6 % and the peak stress is reduced to 35 MPa.

The sudden increase of the contact surface between strands from layer to layer is shown in Figure 18 for a wavelength of 0.007 m. For this wavelength, the contact surface remains constant in the first two layers at the high field region while the contact stress increases linearly towards higher load. In the third layer additional strand-to-strand contacts are formed connected with $L_{w2}$, consequently increasing the contact area. From that layer of, the stress increases with halve the initial slope until further increasing contact area at layer 20, associated with $L_{w3}$ where the process repeats.

All formulations and steps applied in the model, describing the local bending strain and contact stress in each layer are essential for an accurate computation of the local critical current.

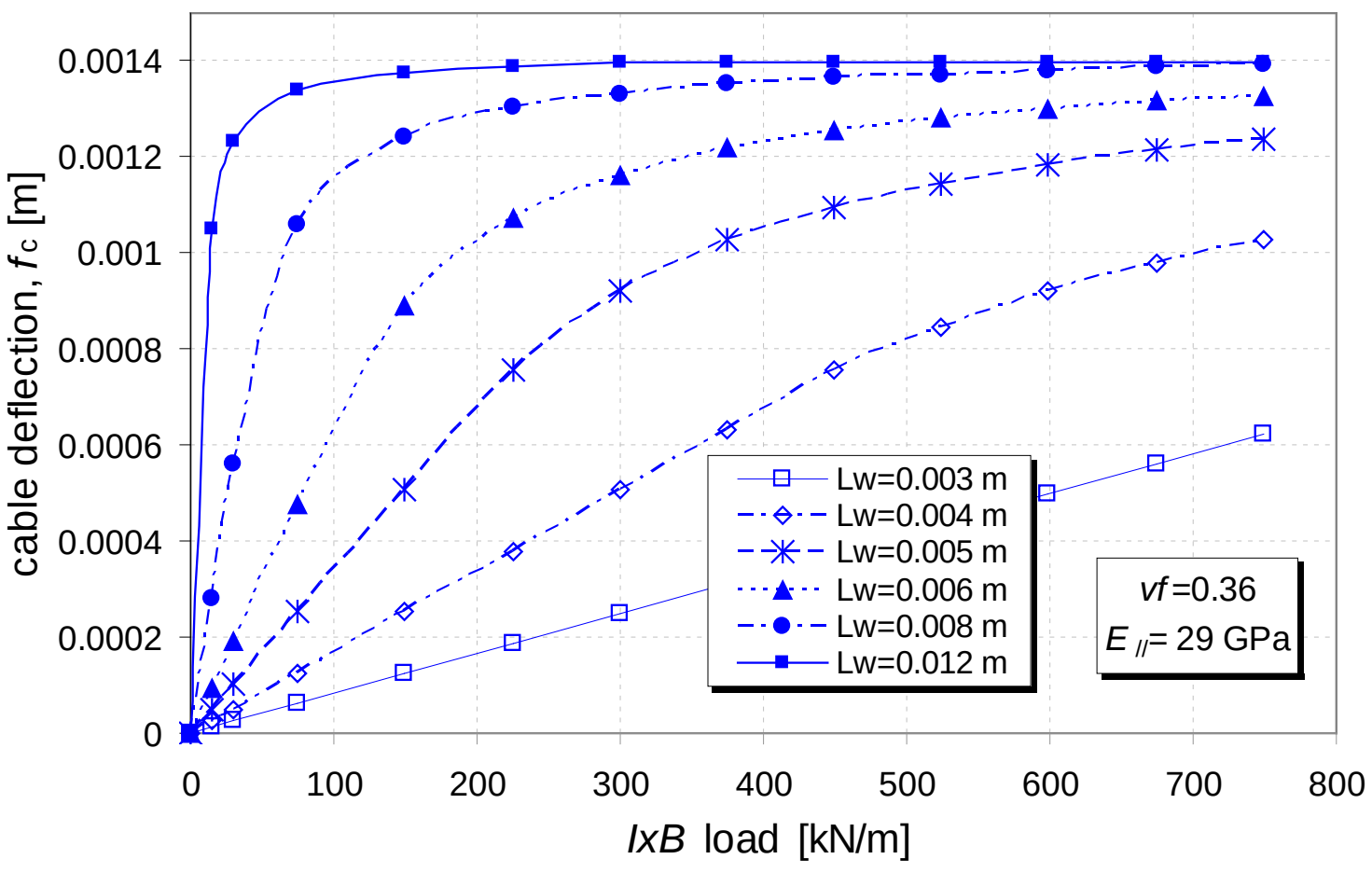

*Figure 12. The cable deformation (compression) versus the applied electromagnetic IxB load for different wavelengths $L_w$ (vf=0.36). The strand properties are from the LMI-EM strand with $E_{||}$ = 29 GPa and $E_{\perp}$ = 3.3 GPa.*

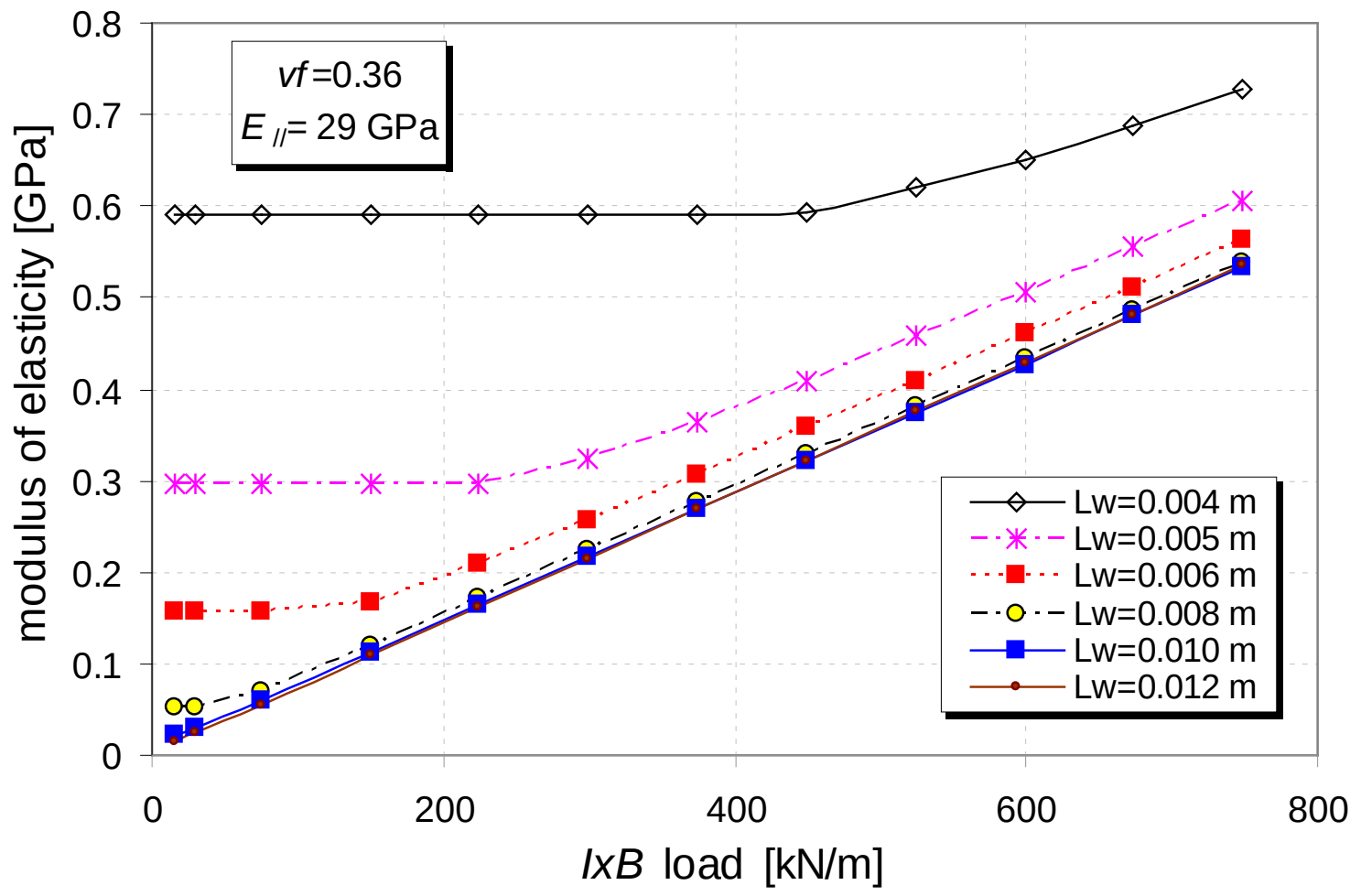

*Figure 13. The overall transverse cable modulus of elasticity versus the applied electromagnetic IxB load for different wavelengths $L_w$ (vf=0.36).*

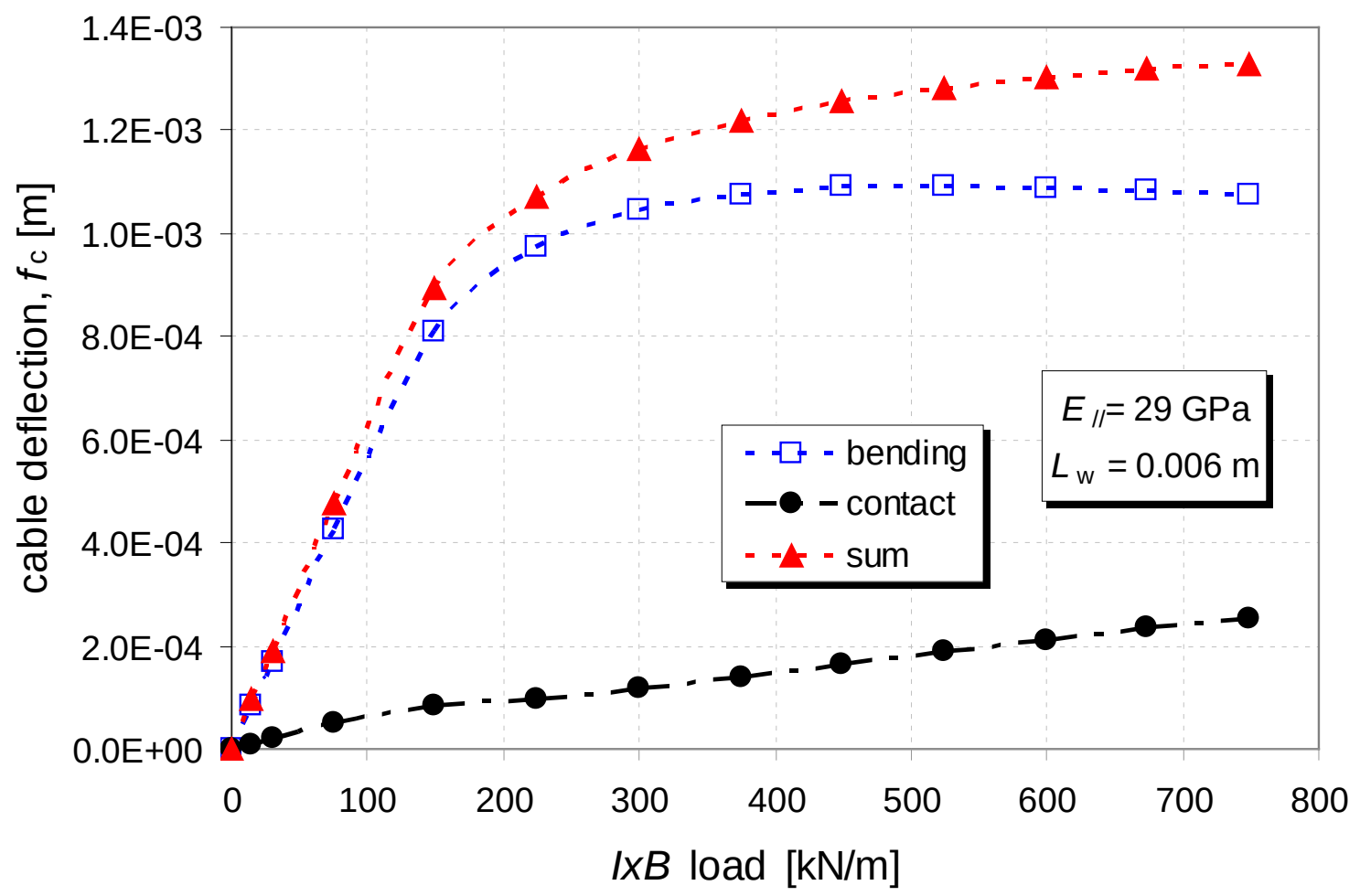

*Figure 14. The transverse cable compression versus the applied electromagnetic (IxB) load for bending, contact stress and both components added (vf=0.36).*

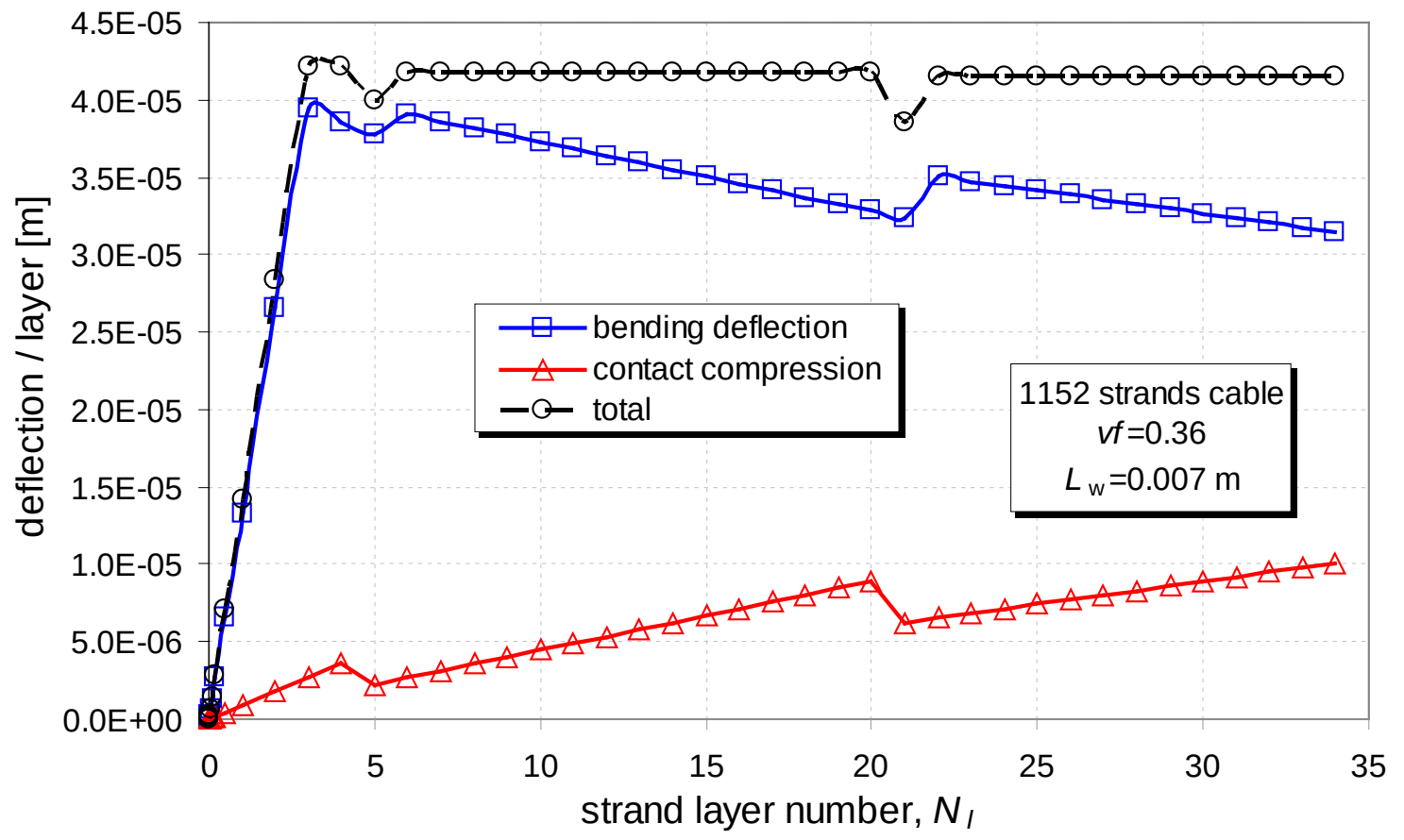

*Figure 15. The bending deflection and strand contact deformation (per strand layer) versus the strand layer number, $N_l$, from high to low field at 750 kN/m load for $L_w$=0.007 m (vf=0.36).*

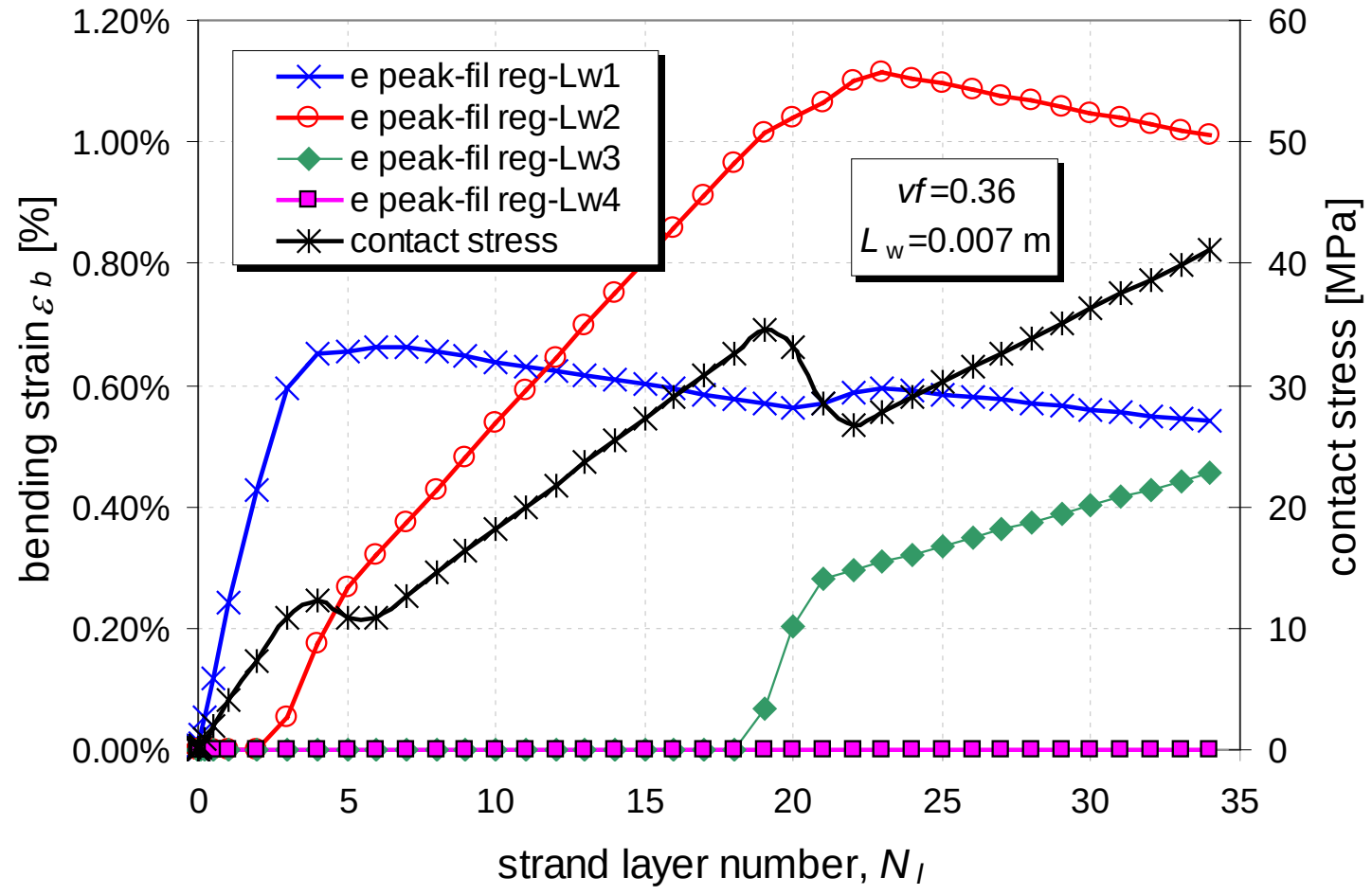

*Figure 16. The bending strain and contact stress (per strand layer) versus the strand layer number, $N_l$, from high to low field at 750 kN/m load for $L_w$=0.007 m (vf=0.36). In the legend the 'e peak-fil reg' is the peak bending strain in the filamentary region or $\varepsilon_{b0}$.*

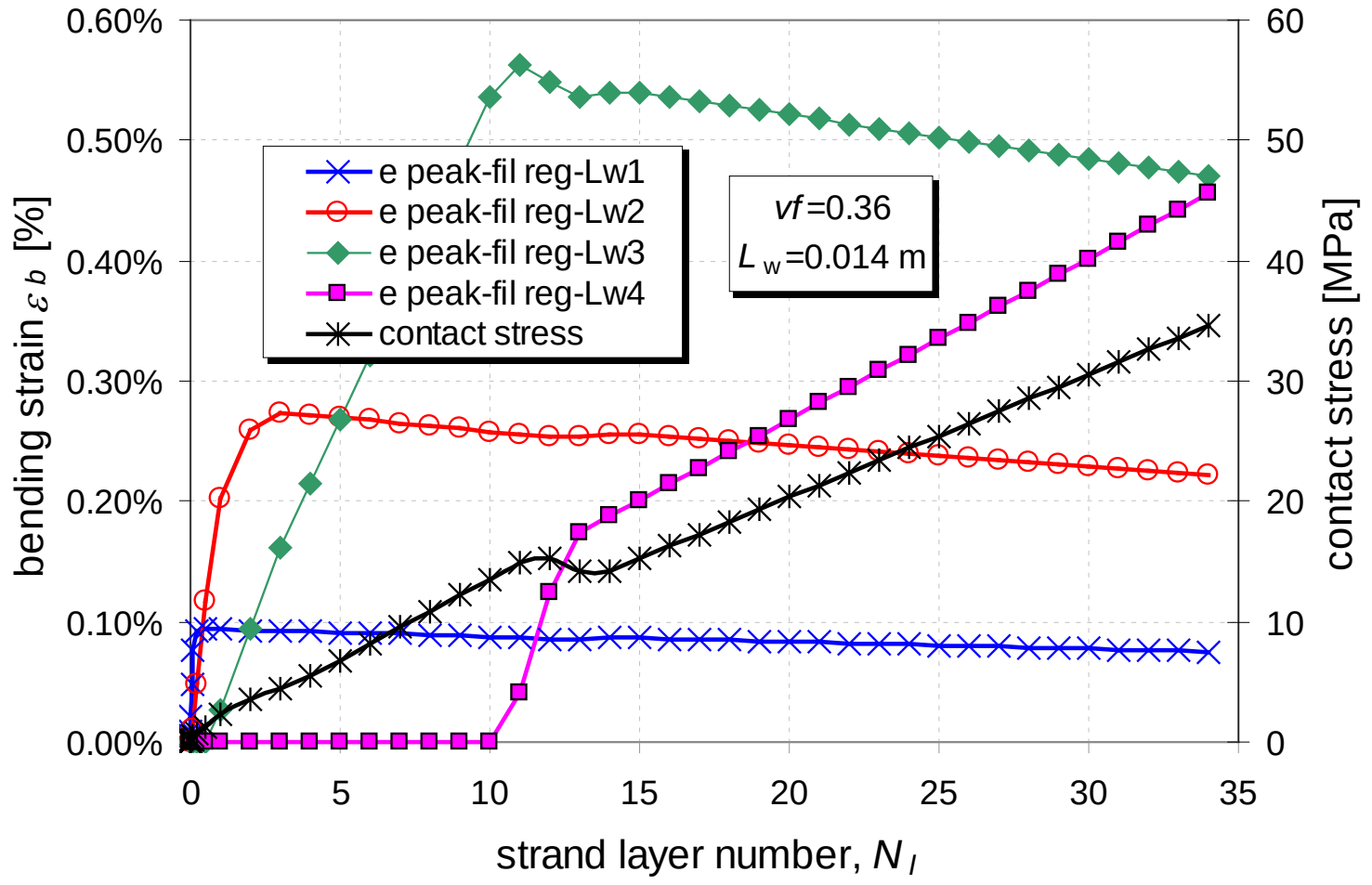

*Figure 17. The bending strain and contact stress (per strand layer) versus the strand layer number, $N_l$, from high to low field at 750 kN/m load for $L_w$=0.014 m (vf=0.36).*

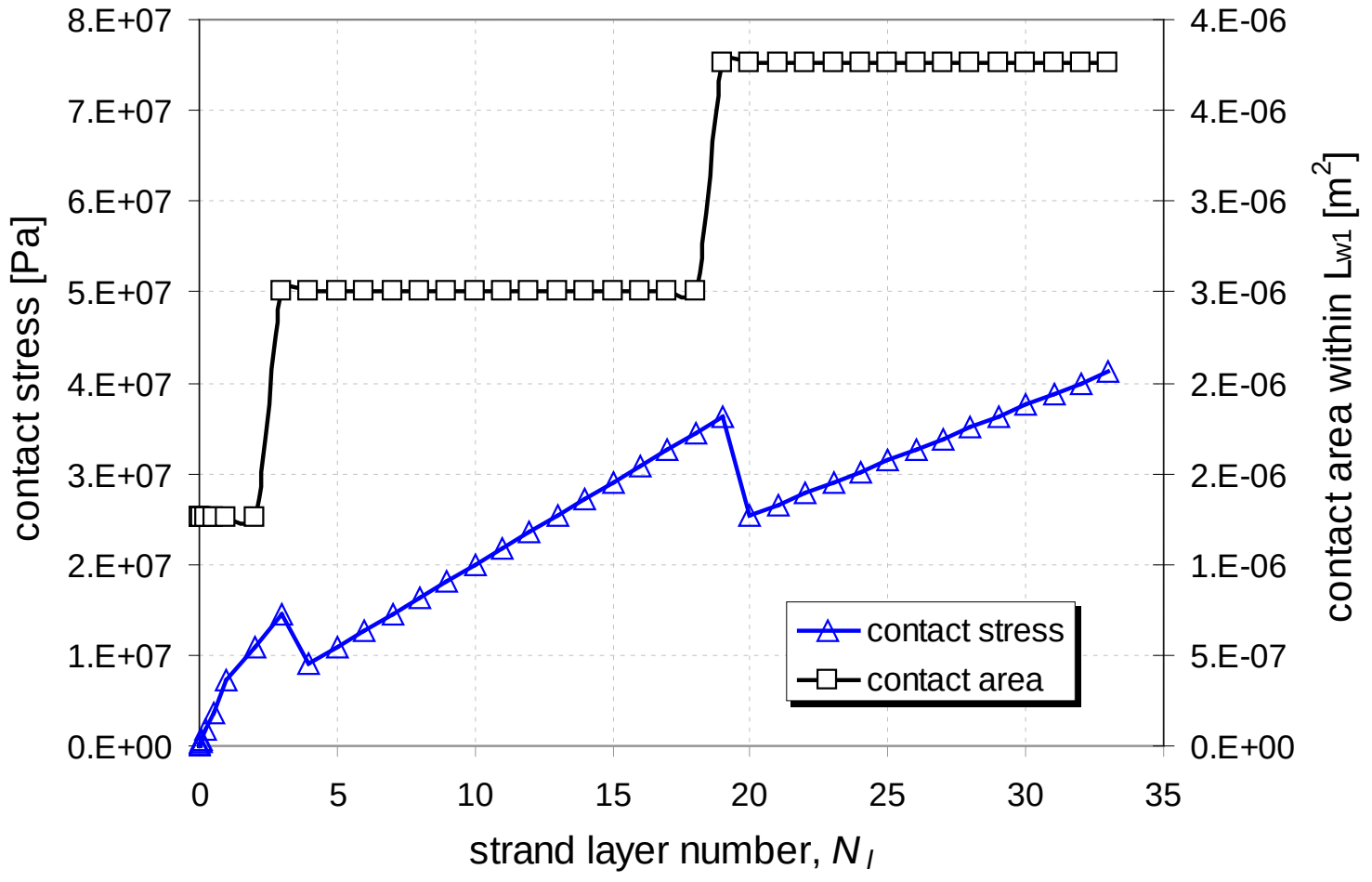

*Figure 18. The interstrand contact stress and the contact area versus the strand layer number, $N_l$, from high to low field at 750 kN/m load for $L_w$=0.007 m (vf=0.36).*

### 5.2. Critical current and transverse *IxB* load

The reduction of the critical current depends on the applied bending and contact load, which varies from layer to layer, and the local magnetic field. Thus, we calculate the $I_c$ reduction for each strand layer number, $N_l$, from high to low field from zero to 750 kN/m load, in a conductor with a void fraction of 0.36 and a wavelength of 0.006 m. The strand properties are from the LMI-EM strand with $E_{//}$=29 GPa and $E_{\perp}$=3.3 GPa. The results for bending and contact load are plotted in Figure 19. The curves represent the reduction from purely interstrand contact load distinguished from the reduction from pure bending throughout the layers of strands in the cable. The reduction from strand contact load is practically nil so almost all reduction is caused by bending. The evolution of the critical current reduction from layer to layer is illustrated in Figure 20 for strand currents ($I_s$) from 5 A to 50 A. A strand current of 50 A corresponds to an *IxB* load of 750 kN/m. In Figure 21 the same is carried out for a wavelength of 0.009 m. The $I_c$ reduction is restricted at the low field (accumulated load) zone of the cable cross-section because the maximum deflection is reached.

In the high field region of the cable cross section, bending is responsible for a rapid decrease of the $I_c$ towards the low field region, until the maximum bending deflection is reached for $L_{w1}$ (Figure 2). The reduction in this high field region is amost completely attributed to bending with the larger wavelength $L_{w1}$ while the reduction related to $L_{w2}$ becomes significant somewhat deeper in the conductor. It appears that when the strand has reached the first bending limit (see Figure 2b) and just before the strand becomes supported by the layer of strands below, ($N_{y+1}$), dividing the wavelength $L_w$ by two ($L_{w1}$ to $L_{w2}$), the bending strain reaches up to 0.8 % for an $L_{w1}$ of 6 mm in layer 5. When the strand becomes actually supported by the strand below, changing $L_{w1}$ to $L_{w2}$, the bending strain reach a maximum of 1.3 % and although this level of strain is extreme, the cable $I_c$ is only moderately affected because this strain occurs in layers $N_y$ with largest accumulated load that are in lowest magnetic field.
When we increase the wavelength $L_{w1}$ to 0.009 m (see Figure 21), we find that the reduction in $I_c$ becomes less excessive, in particular for higher currents. This is due to the lower peak bending strains and contact stresses. The maximum available space for bending leads to lower bending strain for longer $L_w$, while at the same time the strands become supported along longer lengths and on more locations, reducing the local peak contact stress.

When averaging the $I_c$ reduction from all layers for successive *IxB* loads, we obtain the overall cable $I_c$ reduction versus the *IxB* load (relation 22). This is summarised in Figure 22 for different wavelengths. Here we find an impressive improvement in the performance when the wavelength $L_w$ is increased from 6 mm to 20 mm.
The average wavelength ($L_w$) determined from the disassembled full-size ITER CS1 conductor in Figure 4 is 6 mm with a standard deviation of 2 mm. Most wavelengths are thus within the range $L_w$=4 mm and $L_w$=8 mm. When we take the average overall reduction of the $I_c$ versus the *IxB* load calculated for $L_w$= 4, 5, 6, 7, and 8 mm we obtain the upper curve presented in Figure 23. The computed *IxB* behaviour is in good agreement with analyses that has been reported from experiments on coils and short sample tests [3,5]. However, this curve represents the situation for full current transfer. For comparison we show in the same Figure the curve accounting for no current transfer with much stronger degradation. All other results presented here are reflecting full interstrand current transfer.

The remarkable improvement in the performance, when the wavelength $L_w$ is increased from 6 mm to 20 mm, is clearly illustrated in Figure 24 with the overall reduction of the critical current versus the characteristic bending wavelength. It appears that also with a wavelength shorter than 5 mm an improved performance is possible, confirming that the present ITER cable design is unfortunately nearly at the worst possible conductor performance. However, application of shorter twist pitches is not favourable as they lead to an inefficient use of strand and an increase of the cable size. In the same Figure we plotted the peak strain in the strands showing a matching correlation with the $I_c$ reduction. The peak strain and the peak contact stress in the conductor versus the wavelength are presented in Figure 25. Also for the contact stress the same dependency is obtained although there is no effect on the critical current. The influence of the interstrand resistivity, allowing for full current redistribution or non at all, is depicted in Figure 26. It appears that for higher $L_w$ the difference between full and no current transfer becomes less pronounced due to a different balance between the stress and strain concentrations across the cable in relation to the self-field profile. This is well illustrated by Figure 20, showing a small number of layers with severe $I_c$ degradation and Figure 21 with a more balanced distribution of the degradation over the cable cross section.

A decrease of the void fraction also leads to a better performance. This is demonstrated in Figure 27 for void fractions of 0.36 and 0.33, the previous and present ITER conductor design values respectively. The influence of the void fraction is significant and can be explained as limiting the available space for bending deflection.

The effect of the stiffness of the strand is simulated with the properties of the internal tin strand, used for all simulations, and a bronze strand with a lower axial elastic modulus. The result is shown in Figure 28, illustrating that a strand with lower $E_{//}$ is more degraded as it gets stronger deformed by the bending loads.

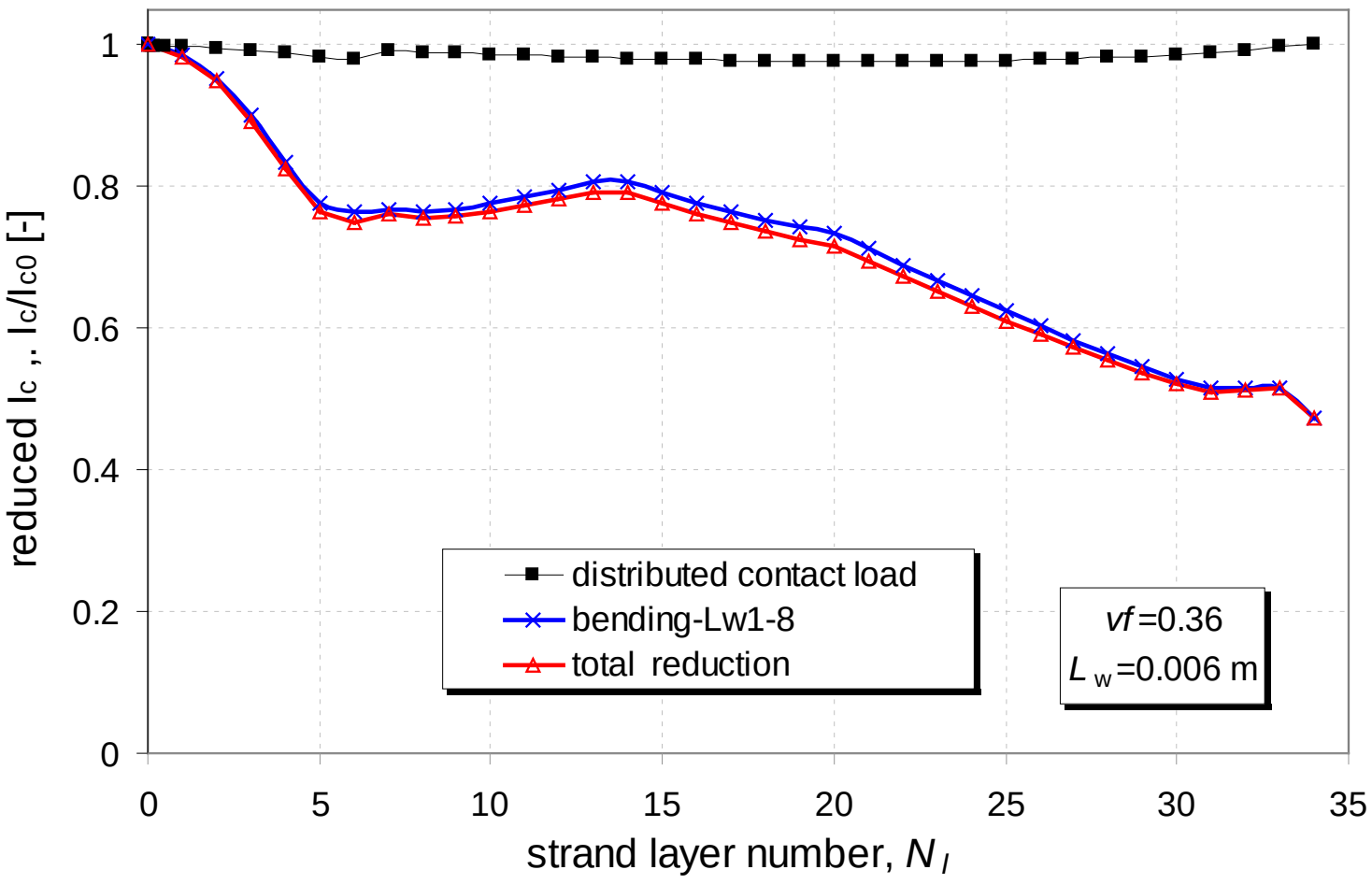

*Figure 19. The reduction of the critical current versus the strand layer number, $N_l$, from high to low field at 750 kN/m load for Lw=0.006 m (vf=0.36). The curves represent the reduction from purely interstrand contact load distinguished from the reduction from pure bending.*

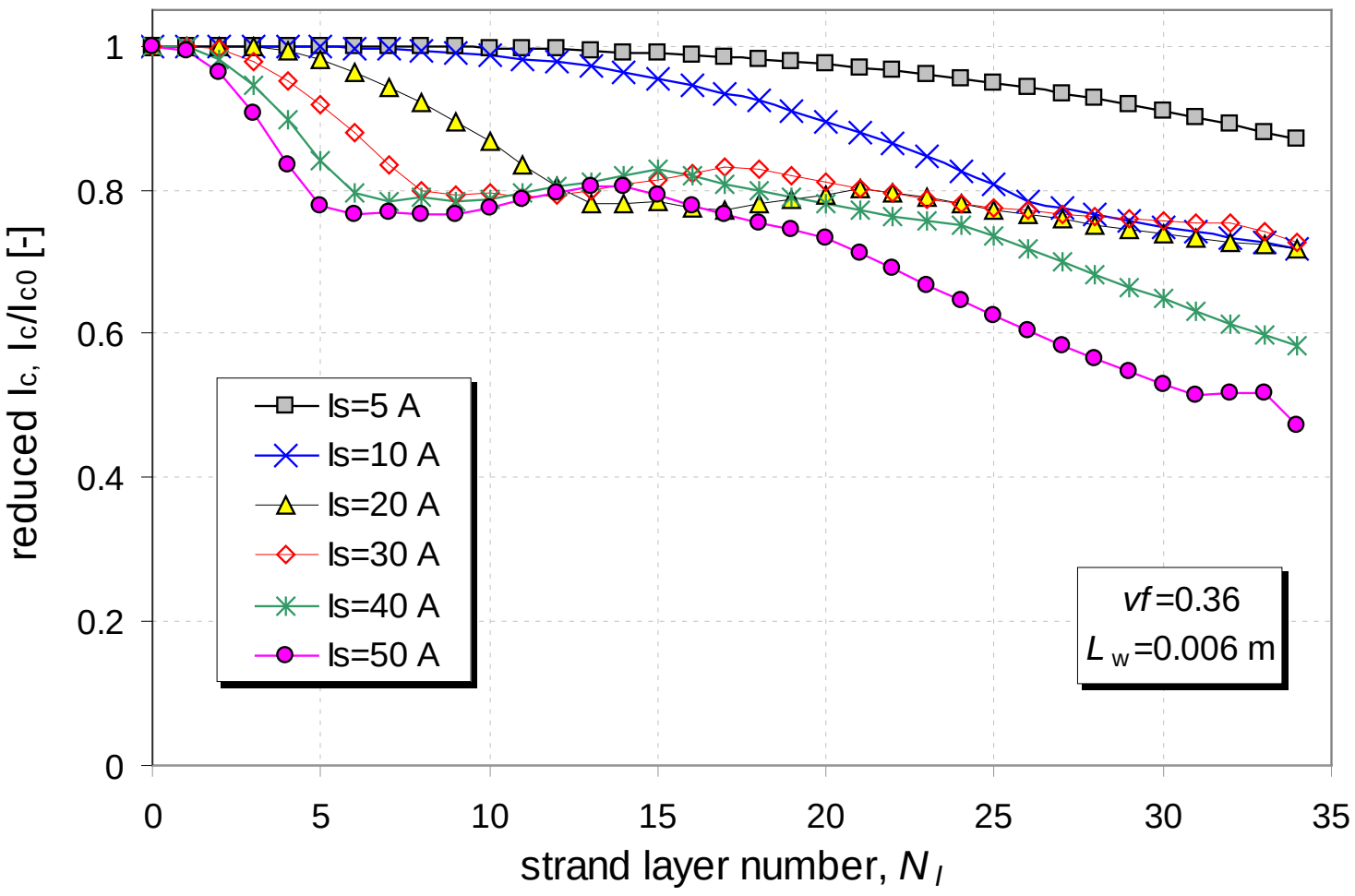

*Figure 20. The reduction of the critical current versus the strand layer number, $N_l$, from high to low field at different strand currents ($I_s$) up to 750 kN/m load for $L_w$=0.006 m (vf=0.36). The curves represent the overall reduction from bending and contact stress.*

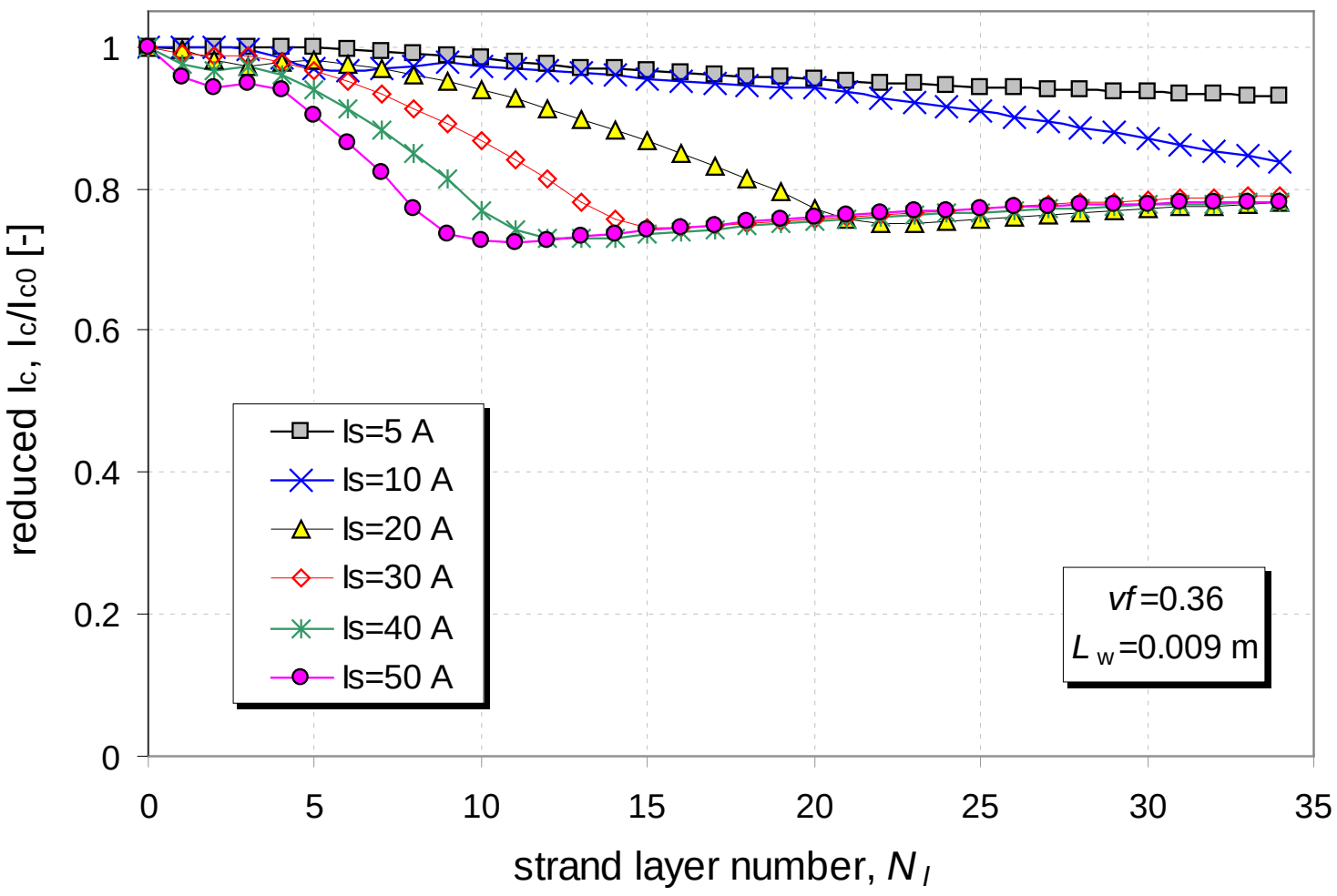

*Figure 21. The reduction of the critical current versus the strand layer number, $N_l$, from high to low field at different strand currents ($I_s$) up to 750 kN/m load for $L_w$=0.009 m (vf=0.36). The curves represent the overall reduction from bending and contact stress.*

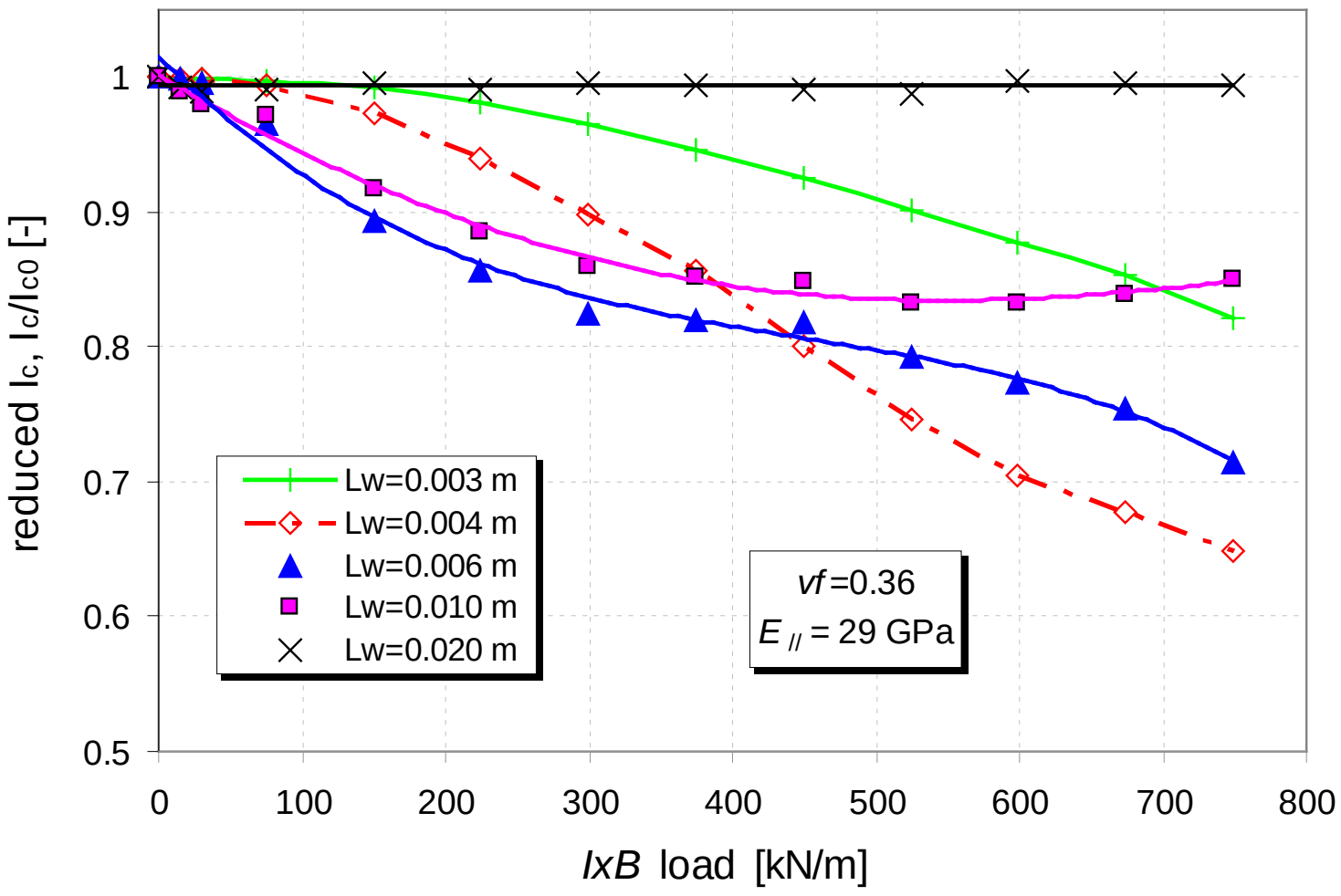

*Figure 22. The overall reduction of the critical current versus the IxB load for different wavelengths from 2 mm to 20 mm (vf=0.36).*

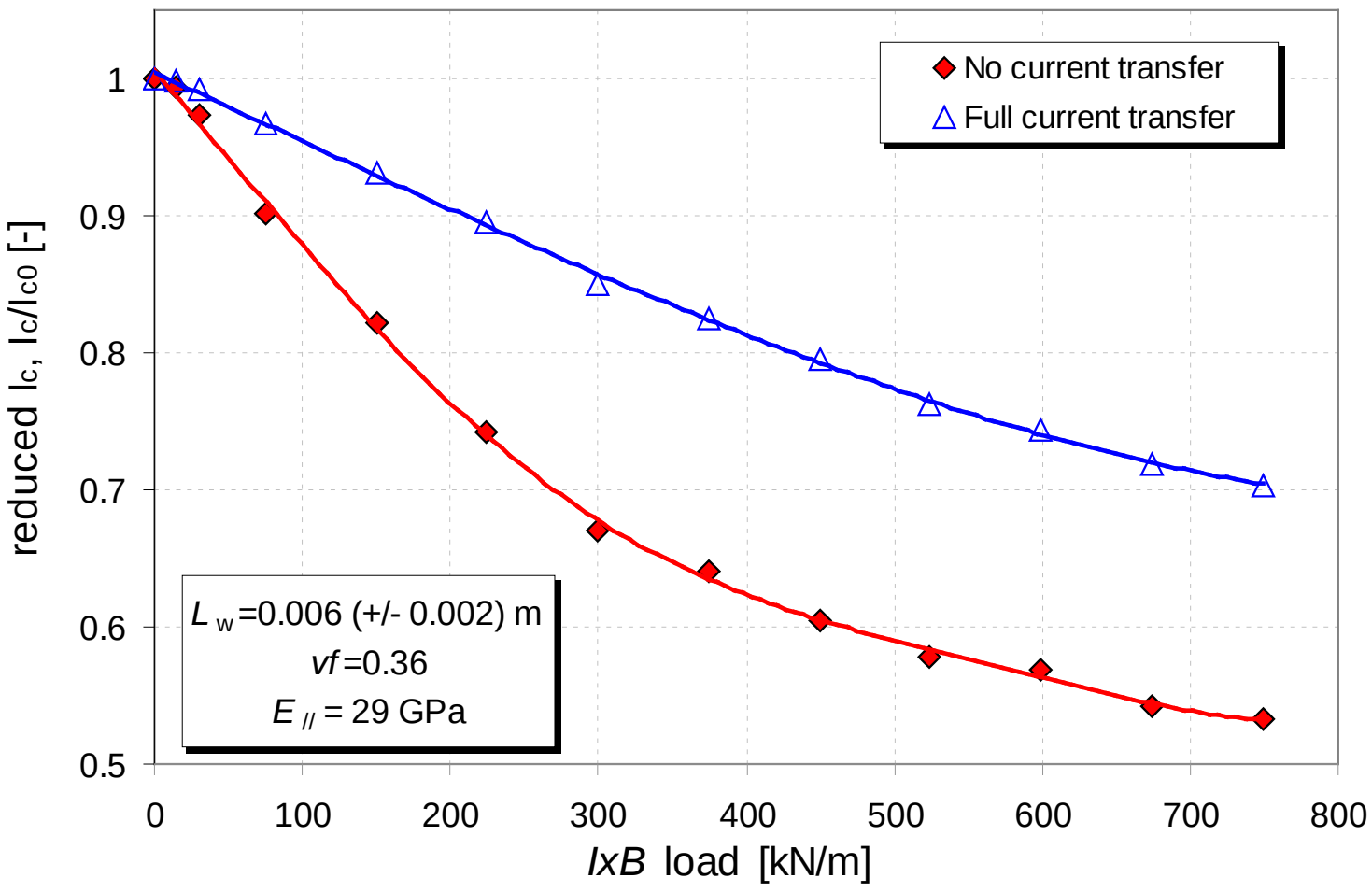

*Figure 23. The computed average overall reduction of the $I_c$ versus the IxB load from $L_w$= 4, 5, 6, 7, and 8 mm representing the spread in the measured $L_w$ from the CS1 conductor with an average $L_w$ of 6 mm and a standard deviation of 2 mm (vf=0.36). The upper curve accounts for full interstrand current transfer while the lower curve represents no current transfer.*

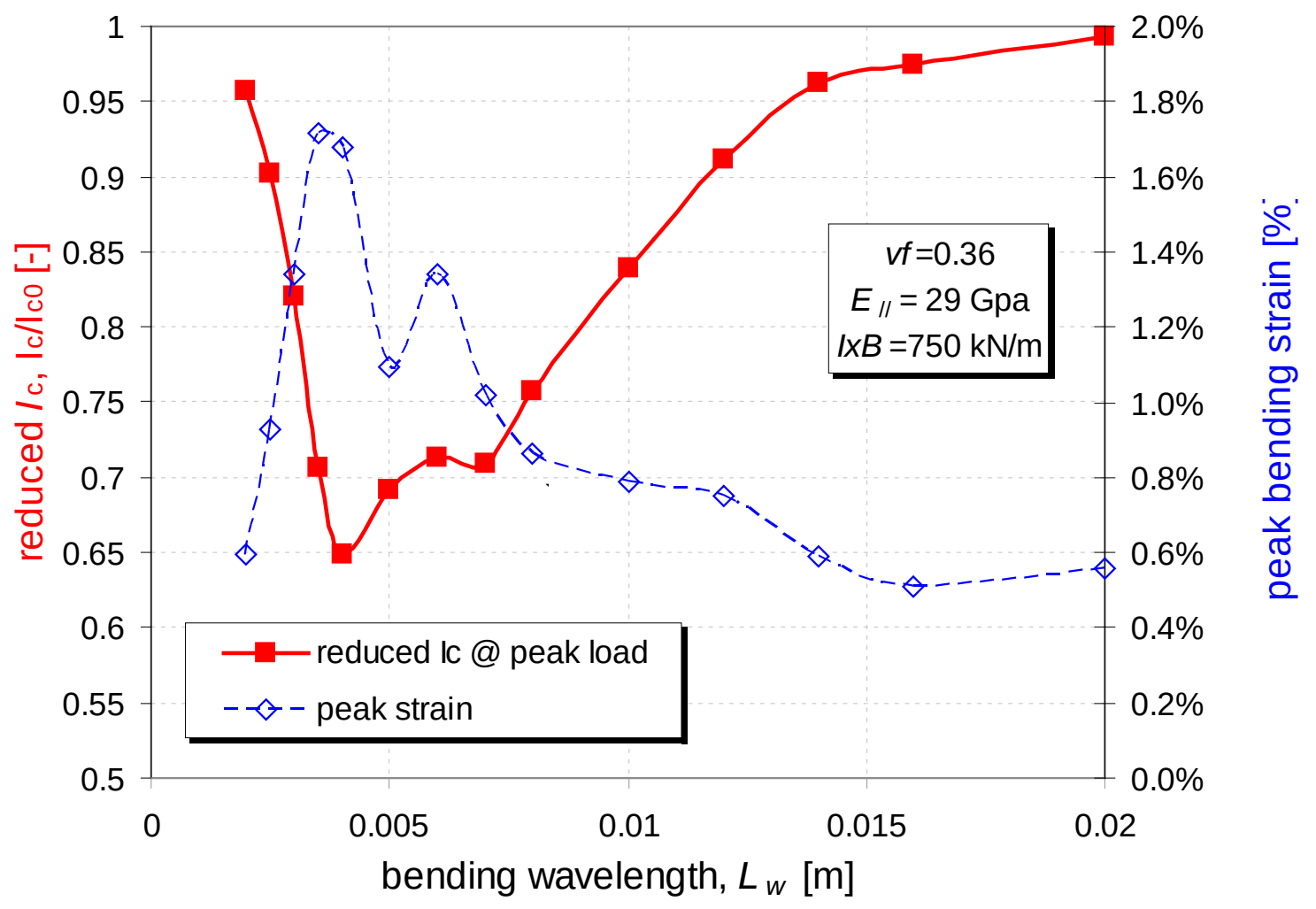

*Figure 24. The overall reduction of the critical current and the peak bending strain in the cable cross – section versus the characteristic bending wavelength (vf=0.36).*

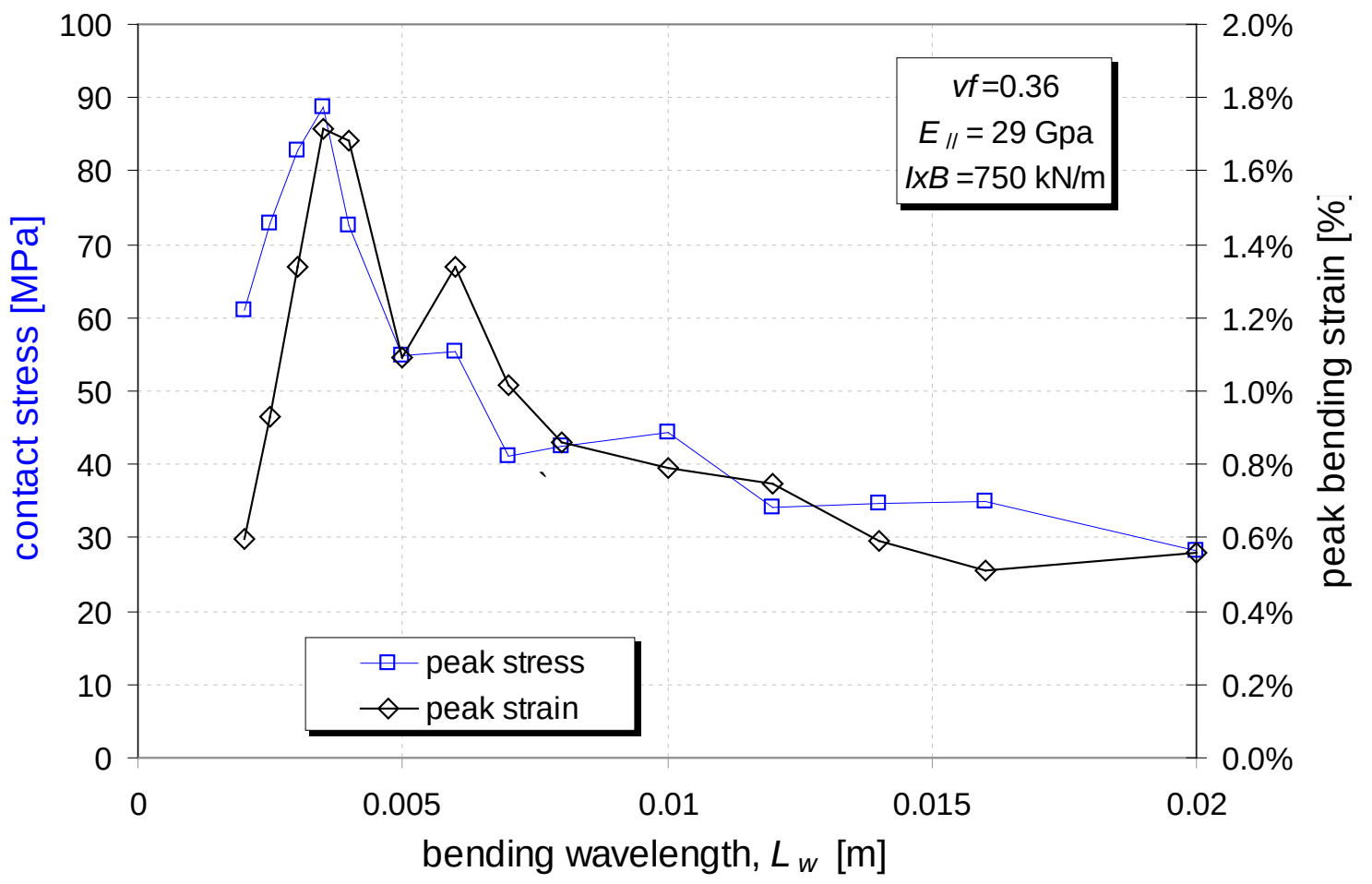

*Figure 25. The peak contact stress and peak bending strain in the cable cross section versus the characteristic bending wavelength (vf=0.36).*

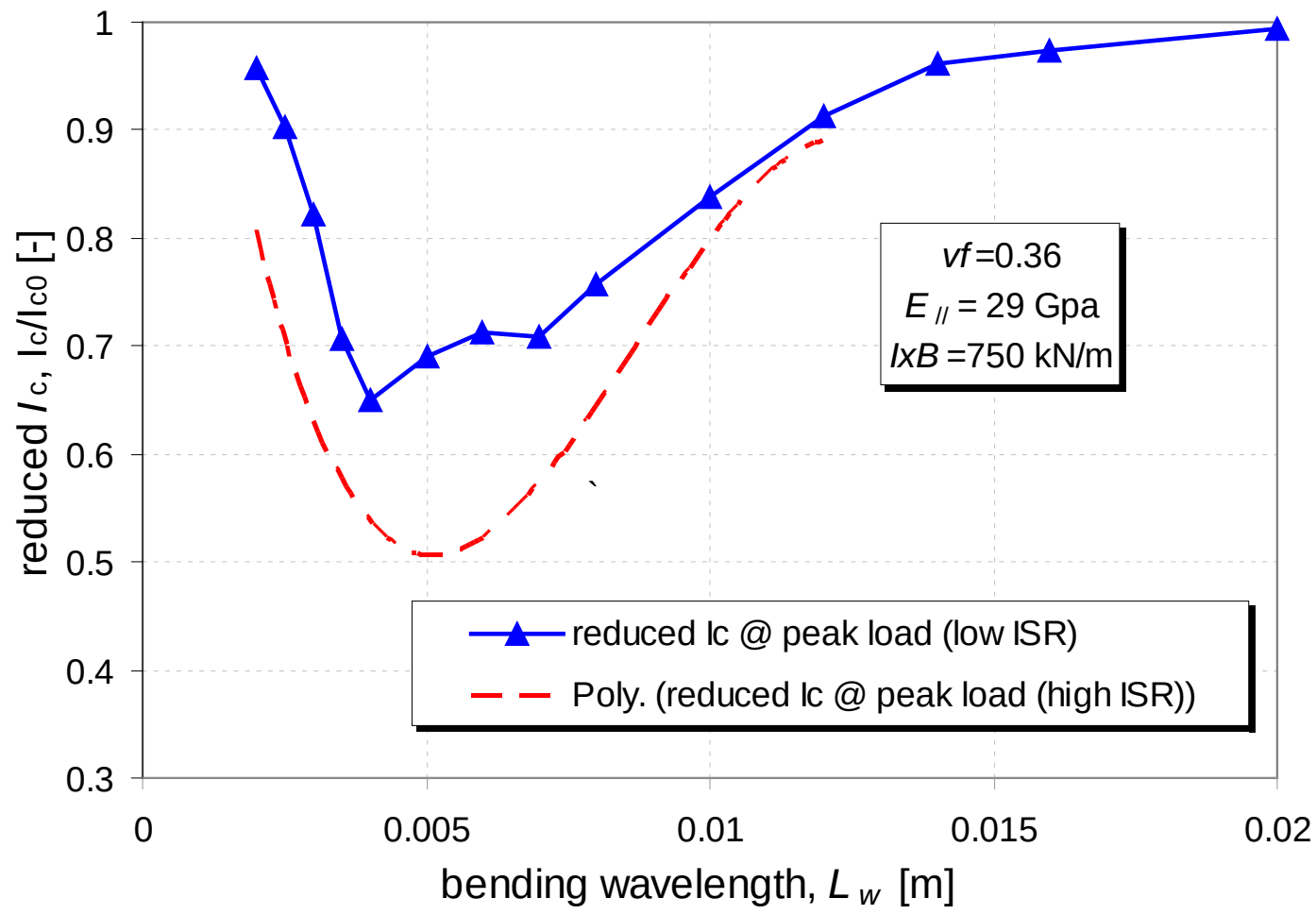

*Figure 26. The overall reduction of the critical current for full interstrand current transfer (low interstrand resistance, ISR) and no current transfer (high ISR, polynomial fit) in the cable cross –section versus the characteristic bending wavelength (vf=0.36).*

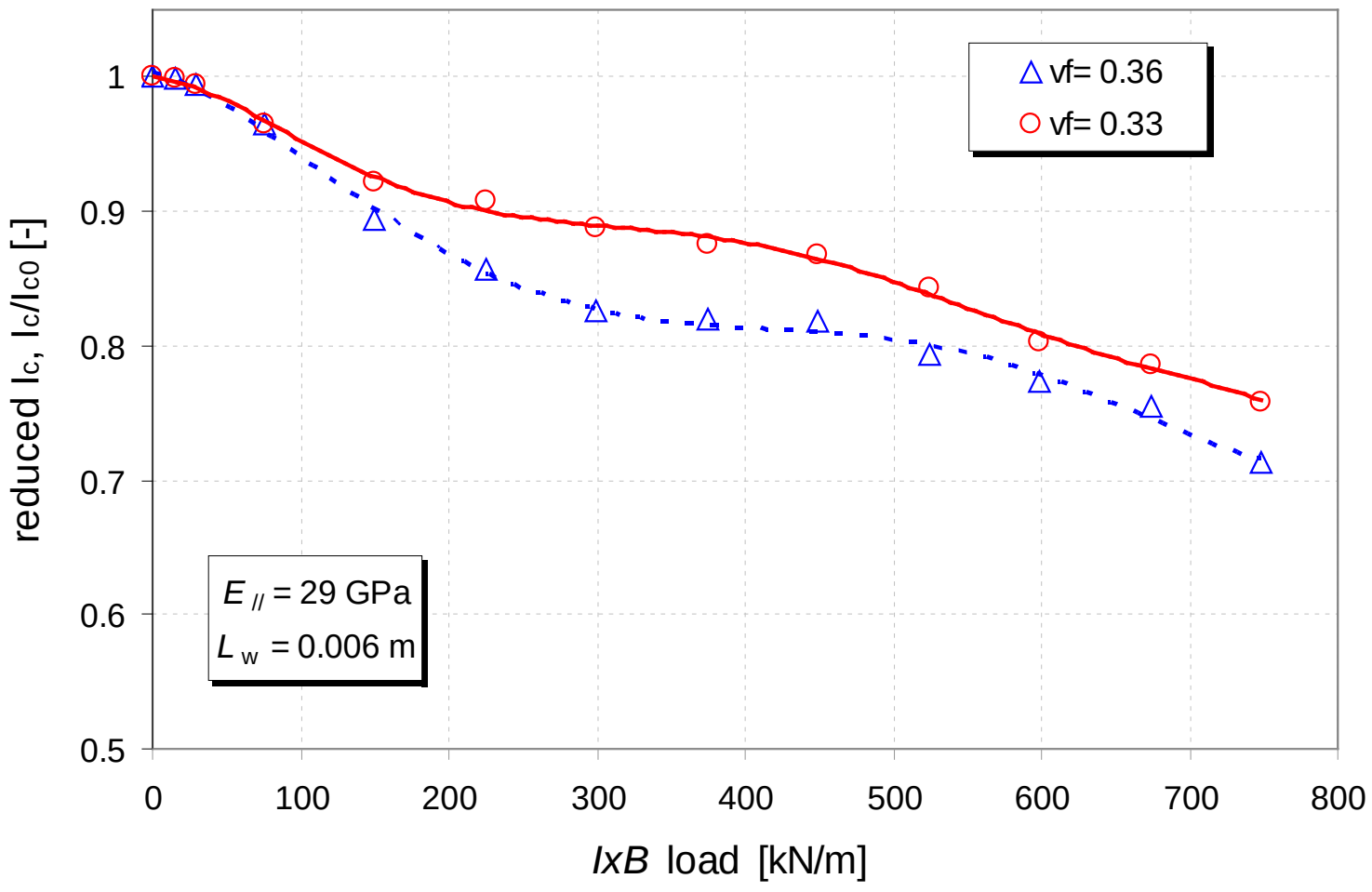

*Figure 27. The overall reduction of the critical current versus the IxB load for conductors with a void fraction of 0.36 and 0.33.*

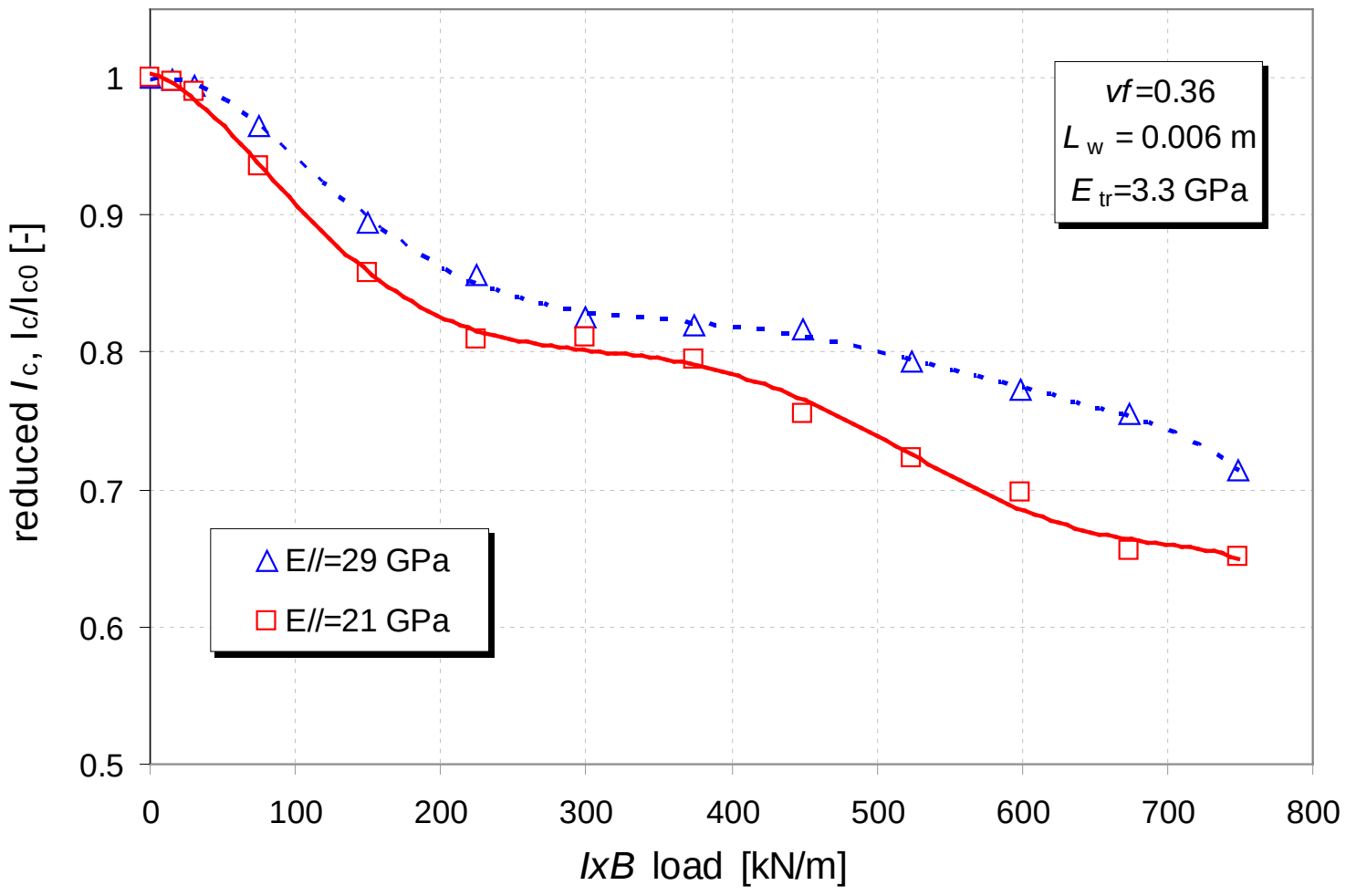

*Figure 28. The overall reduction of the critical current versus the IxB load (vf=0.36). The strand properties are from the LMI-EM internal tin strand with $E_{||}$ = 29 GPa and $E_{\perp}$ = 3.3 GPa and the VAC bronze strand wit. $E_{||}$ = 21 GPa and $E_{\perp}$ = 3.0 GPa*

### 5.3 Evolution of peak strain with $L_{w(n)}$

The relation between the local peak strain and stress is depicted in Figure 24 and Figure 25. However, the local peak strain and the number of higher orders $n$, required to describe the entire cable ($N_l$) strongly depend on the wavelength $L_{w1}$. Knowing now that for longer wavelength the peak strain and stress relaxes, we explore how the peak strain evolves for the higher order wavelengths. The extension to higher order bending wavelengths in the direction of load accumulation from layer to layer, successively going from $L_{w(n)}$ to $L_{w(n+1)}$, is shown in Figure 29. The peak strain throughout the cable decreases, when higher order wavelengths become involved. For extremely short wavelengths, the higher orders for $n$ do not occur in the cable bundle and $n$=1. With increasing $L_{w1}$, the peak strain increases until the higher orders of $n$ become involved. For longer $L_w$, $n$ becomes higher together with the homogenisation of the contact stress and bending strain peaks.

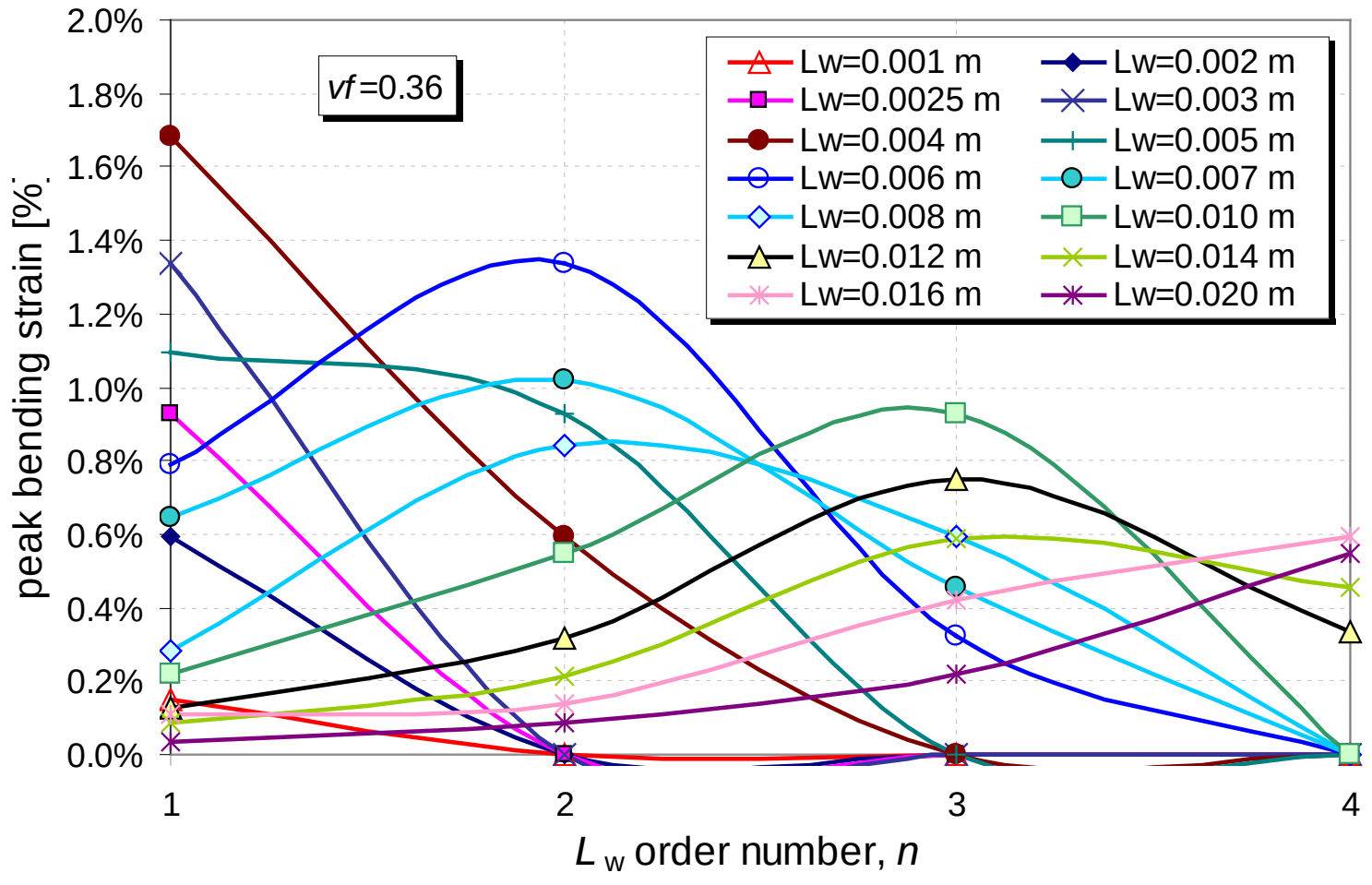

*Figure 29. Computed peak bending strain versus the higher order number n of the relevant wavelength $L_{w1}$ for different characteristic bending wavelength, $L_{w1}$ of a cable.*

## 6 Discussion

### 6.1. Mechanical response of the cable

The TEMLOP-Model predicts a total cable compression of 1330 µm, with 1070 µm for bending and 250 µm for direct strand contact stress, in a cable with a void fraction of 0.36 and a characteristic wavelength of 6 mm. This is in line with the experimental values obtained from the Cryogenic Press [25] and is obviously connected to the derived limitation in bending deflection $f_{sb}$ from relation 9 and the strand contact deformation adopted from Figure 8. The experimental values found for the cable compression and correlated displacement per strand, for three conductors with different void fraction, are plotted in Figure 30 [36]. For the CS1 type of cable with a void fraction of 0.36 we found a maximum deflection ($f_{cm}$) of 1040 µm [25]. The measured compression per strand then reaches a maximum amplitude of 34 µm. At this point we can remark that also from thermo hydraulic analysis, it was deduced that a so-called third channel between one side of the cable and the inner conduit wall, due to cable bundle deformation after cyclic loading of the CS Insert Model Coil, would entail a gap of about 1 mm [37].

For a cable with a void fraction of 0.33, the model predicts a total cable compression of 1040 µm, with 810 µm for bending and 225 µm for direct strand contact stress ($L_w$ of 6 mm). This is in agreement with what is expected by interpolation for the total deflection in Figure 30. However, the influence of the void fraction on the cable compression as computed with the model is stronger than obtained from the press measurements. This may be due to an underestimation of the total contact deformation as derived from Figure 8 and the assumption that the sum of the bending and contact deformation reflects the total compression, which is not necessarily conform the reality.

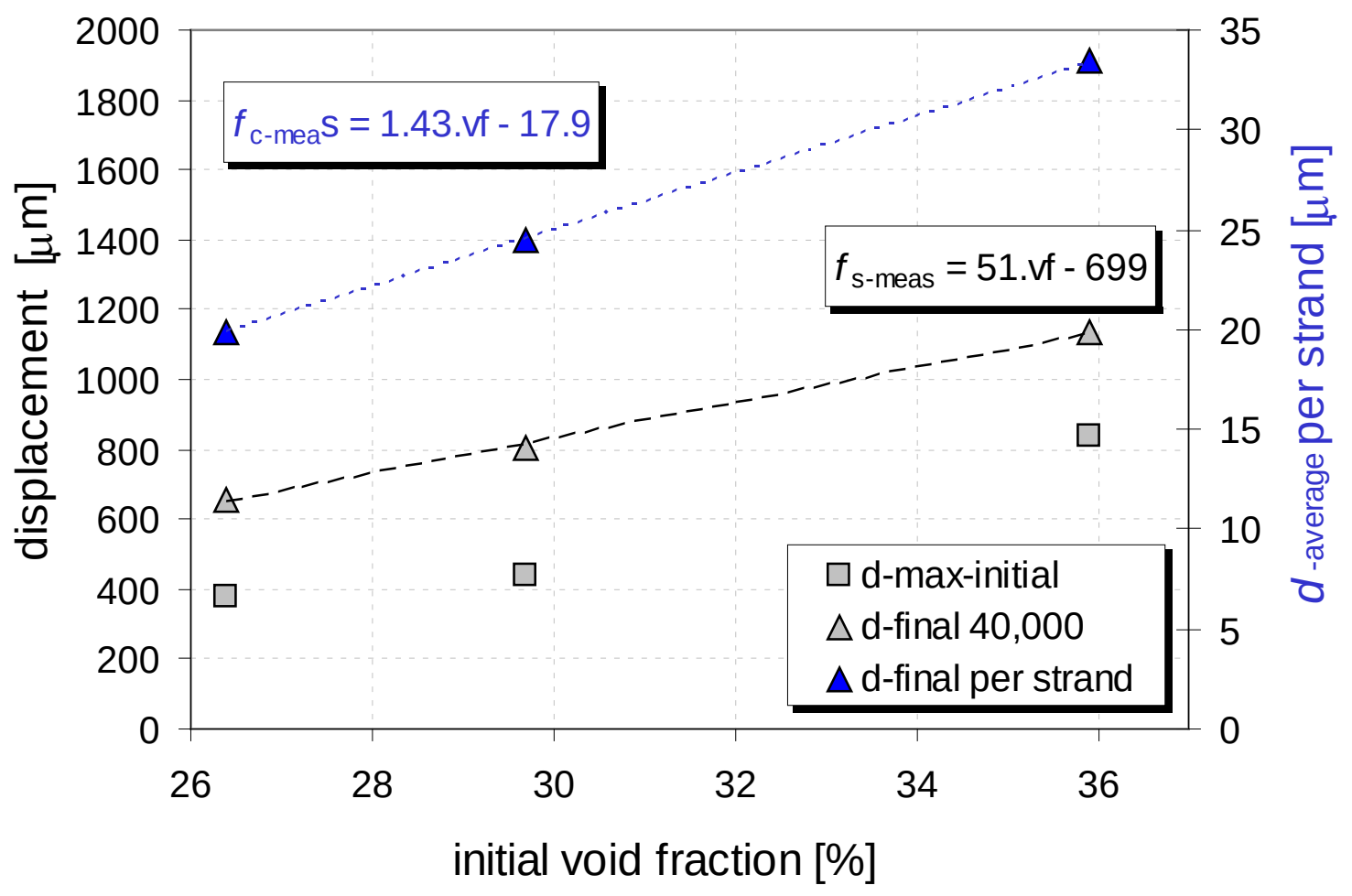

*Figure 30. The total compression of several Model Coil conductors with different void fraction and the deduced possible free displacement for bending per strand.*

Transverse stress-strain measurements performed with the TARSIS crossing-strands probe on the LMI-EM strand showed that the loading curve is practically without hysteresis when unloading (see Figure 8). The low transverse modulus of elasticity of the strands, compared to the modulus in axial direction being one order of magnitude higher, implicitly suggests that the transverse cable bundle compression through contact stress is connected to mostly plastic deformation of strand crossings and line contacts. The plastic deformation from transverse interstrand contact deformation is primarily determined by the copper stabiliser, practically immediately yielding at 4.2 K, while the axial strand stiffness, linked to bending, is significantly enhanced by the $Nb_3Sn$ layer [18]. The conductor cross section depicted in Figure 7 illustrates the compaction of a cable after test in the Twente Cable Press with peak transverse load for 40,000 cycles. At the upper perimeter of the cable bundle, a permanent plastic deformation of 0.6 mm is clearly visible leaving a gap between cable and inner conduit wall. This 0.6 mm plastic deformation consists of 250 μm representing direct strand contact stress deformation according to the model and the remaining part can be attributed to plastic deformation from bending. This leaves about 0.5 mm scope for compression with elastic bending after cycling of the conductor. The so-called *IxB* dependency, as it is often presented in a $\varepsilon_{extra}(IxB)$ plot, is then directly linked to the remaining span for elastic bending. The $\varepsilon_{extra}$ represents the additional axial strain that is required to match the measured strand performance to the degraded cable performance [5].
As the model is not adapted to comply for plastic deformation and creep, the performance reduction with cycling of the load is not reflected in the results.

The overall modulus of elasticity measured on a CS1 conductor in the Twente Cryogenic Press is plot in Figure 31 together with the computed curves for $L_w$=6 mm, showing good agreement.

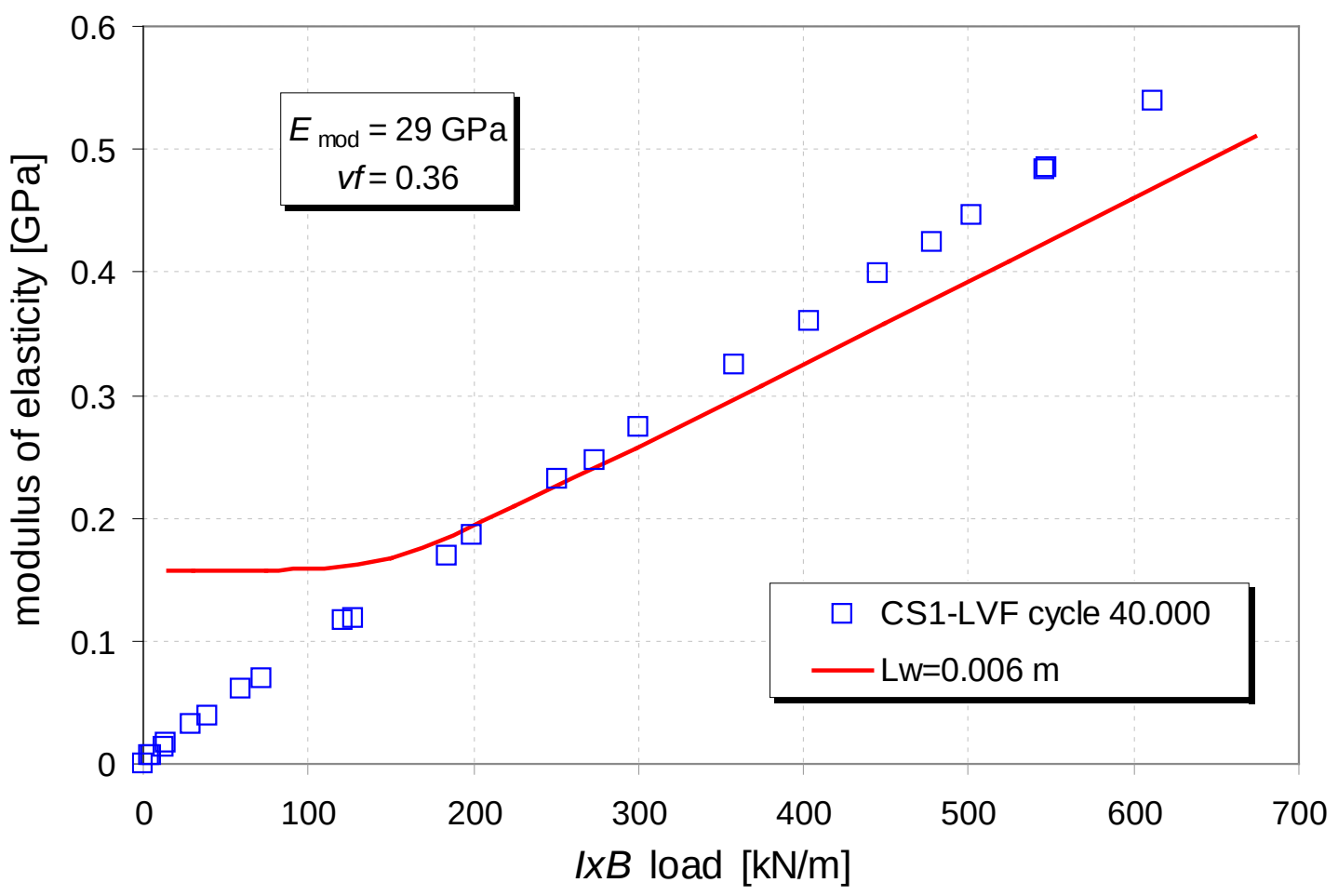

*Figure 31. The measured and computed overall transverse cable modulus of elasticity versus the applied electromagnetic IxB load for the CS1 conductor (vf=0.36). The strand properties are from the LMI-EM strand with $E_{||}$ = 29 GPa and $E_{\perp}$ = 3.3 GPa.*

If we compare the computed deflection of the cable versus the applied load for different wavelengths to the measured cable compression in the Twente Press (see Figure 32), we find that the behaviour of the cable for increasing number of cycles in the experiment, could be described by the model as for increasing $L_w$. The model obviously does not incorporate the effect of plastic deformation and work hardening with cycling, but there is a strong analogy in the characteristics. The computed mechanical response with increasing $L_w$ is linked to a decreasing influence of bending at the same time accompanied with a growing contact surface. In that sense we can understand the measured cable deformation, during already the first cycle, as largely plastic deformation. With further cycling of the load on the cable the component of the plastic deformation decays finally leaving only the component of mainly elastic bending.

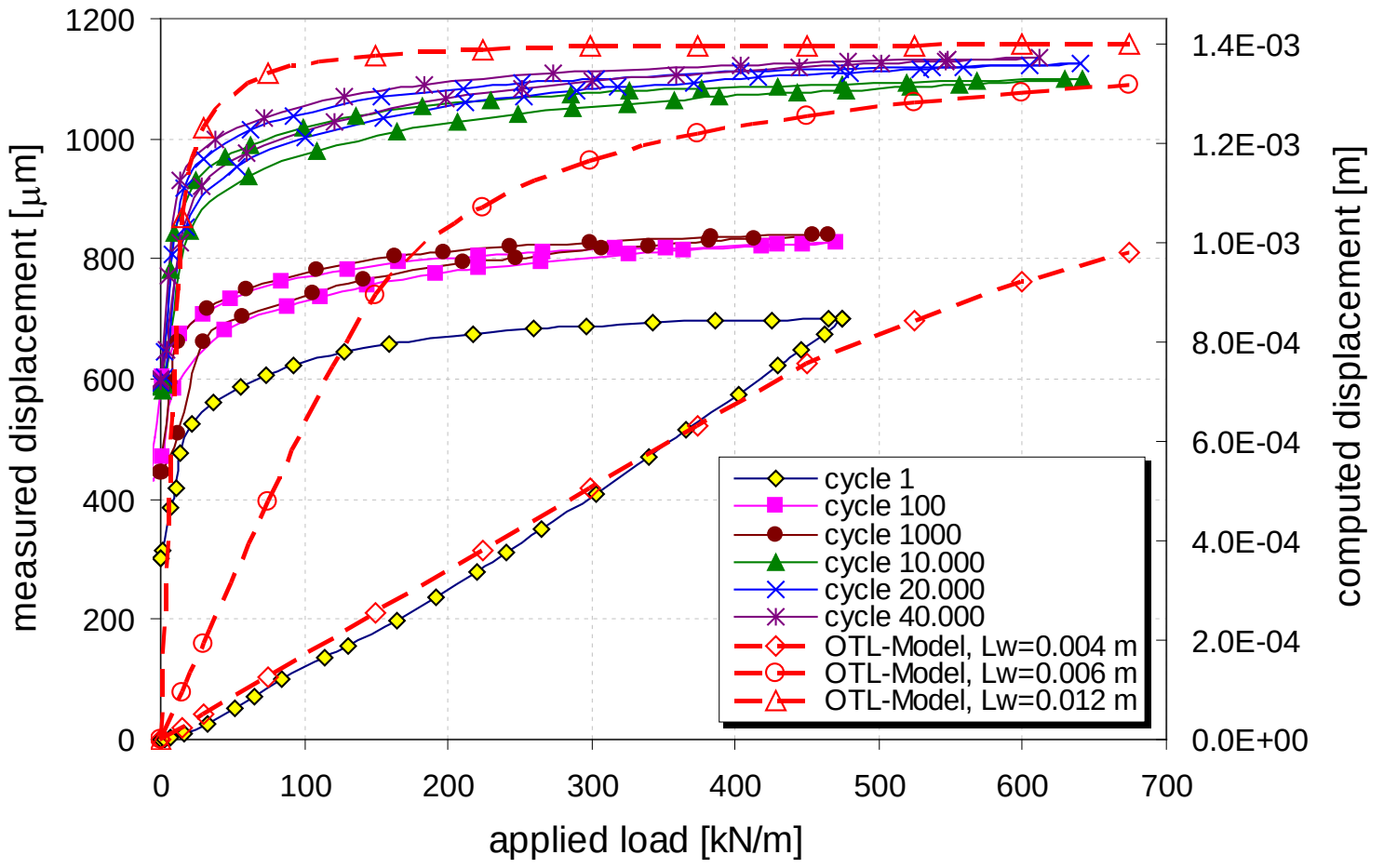

*Figure 32. The measured and computed overall transverse cable compression versus the applied electromagnetic IxB load (vf=0.36). The experimental results are from the CS1 conductor along cyclic loading while the computed curves are for different $L_w$. The strand properties for the computation are from the LMI-EM strand with $E_{||}$ = 29 GPa and $E_{\perp}$ = 3.3 GPa.*

## 6.2. Critical current reduction

One of the main aims of the TEMLOP-Model is to verify the possible influence of the cabling pitches on the performance degradation originated by transverse load. As already put forward in [25], the most favourable transverse load distribution is a homogeneous one, representing an almost infinite line contact of parallel strands. The most important prediction of the model, in terms of optimisation the conductor performance by increasing the twist pitches, is understood as a homogenisation of the load distribution or in other words, a significant lowering of the local peak strains. This is clearly illustrated in Figure 24 and Figure 25.

The outcome of the model can be explained by exploring the limiting cases for $L_w$. When $L_w \rightarrow \infty$, the angle between the crossing strands goes to zero and the stress reaches to the homogeneously distributed case represented by relation 4. At the same time the bending diminishes as the strands become in fact parallel resulting into minimisation of the bending strain. For $L_w \rightarrow 0$, the bending strain also goes to zero as reflected by relation 2 and the contact stress reduces to the homogeneity limit from relation 4 with contact area going to the maximum. With $L_w$ in between the limiting cases, periodic bending strain and local concentration of contact stresses lead to reduction of $I_c$.

Moreover, it is not difficult to imagine that an increase of the twist pitches advances the flexibility in the self-regulating mechanism of strands being forced to find the mechanically most favourable position in the cable bundle. This way the strands find better distributed support against bending from subsequent layers of strands towards the accumulated load zone in the cable cross section.

In Figure 4 it is shown that for the CS1 type of conductor, the periodicity in the number of strand crossings is relatively high with a wavelength of 6 mm. The first cabling stage triplet has a twist pitch of 45 mm and the second stage quadruplet has a pitch of 74 mm. In the CS2 type of conductor from Figure 5, with much longer twist pitches (about 70 mm and 90 mm respectively) we find a less frequent periodicity with an average wavelength larger than 6 mm. At the same time, we observe a much smaller crossing angle than the average 45 degrees from the CS1 type of conductor.

In view of the model predictions, concerning the dependency between the twist pitch and transverse load distribution along the strands in the conductor, the cabling scheme of the CS2 seems much more advantageous for large conductors with high transverse load [25]. Unfortunately, this type of conductor was only in the low field region of the CSMC and was not examined on its temperature margin. A test of the CSMC outer module (without inner module) under appropriate conditions to explore the CS2 performance could serve as a reference test for the prediction of the TEMLOP-Model.

Unfortunately the cabling scheme for CICCs and not only for ITER conductors, from which experimental data are available, is kept practically the same along the years. As said, this is likely the main reason that the influence of cabling remained unknown.

In the Introduction we mention that part of the analysts in the fusion community postulate that a significant fraction of the $Nb_3Sn$ layer in the strand material of the cable is breaking [3,10] but that there is no conclusive evidence. The possibility of filament fracture is supported by experimental work [23] even though no results of an internal inspection of strands from a tested conductor have been presented yet. However, although it is not relevant for the analysis with TEMLOP presented here, whether the permanent degradation is due to filament breakage or plastic deformation of strands, the outcome of the model gives an indication. At the experiments on the TARSIS bending probe with stepwise increase of the load with intermediate unloading, we find a clear irreversible degradation of the $I_c$ and the *n*-value [1]. In general we observe a permanent degradation in $I_c$ when exceeding an applied bending strain of 0.3 ÷ 0.4 % with the TARSIS probe. At the same time we find a permanent deformation of the strand as for zero load the strand deflection never returns to zero after having applied a certain load level. This means that from only this experiment we are not able to distinguish between plastic deformation and filament breakage. The irreversible strain limit of about 0.4 % corresponds more or less to the applied tensile strain that is required for irreversible degradation of the $I_c$ in experiments for strand axial strain testing [28]. With TEMLOP we find an applied peak bending strain in the $Nb_3Sn$ filaments of about 1.3 % for the presently used cabling scheme (Figure 24, Figure 25 and Figure 29). Assuming that a steel conduit would impose an additional axial pre-compression leading to roughly -0.7 % intrinsic axial strain in the filaments after cool-down, the bending strain would lead to a net tensile intrinsic peak strain of + 0.6 %. If in addition, we also take into account that at the location with peak bending strain the contact stress reaches maxima as well, local filament breakage seems evident.

### 6.4. Related effects

The most favourable transverse load distribution is obviously a homogeneous one, representing an almost infinite line contact of parallel strands [25]. That pleads for long cabling pitch lengths but this is in conflict with the restrictions given by the generation of coupling loss. The AC loss in terms of interstrand coupling loss is the driving factor for minimization of the cabling pitch lengths together with compelling adequate stiffness to the cable bundle. In principle the generation of coupling loss increases with the square of the pitch length as long as the contact resistance remains constant. In that sense the increase of the coupling loss is a potential disadvantage when choosing longer pitches to enhance transverse load optimisation. However, for the CS2 type of cabling structure we find a coupling loss time constant in the virgin state of about 100 ms, while for the CS1 type a value twice as high is obtained [38]. We take note that a third part of the strands in the CS2 is copper strand instead of all superconducting wire as in the CS1-type. In general, we observe that the coupling loss is roughly proportional to the amount of superconducting strands for comparable cable geometries. From this we can expect that an extrapolation of the coupling loss by only using the ratio of the twist pitch is by far too conservative. It is easy to understand that with increasing the cabling pitches, the contact surface between strands expands while simultaneously the contact stress reduces. An increase of the contact resistance may be a likely explanation for the relatively low level of coupling loss from the CS2-type of conductor, counterbalancing the effect of the increased cable pitches [39]. Consequently, a serious drawback from a possible increase of the coupling loss is not expected.

The transparency of the cable bundle for the cooling medium (liquid helium under pressure) might be slightly affected but also here we do not expect a noticeable impact with a nominal void fraction of 0.33.

An advantage of longer cabling pitches besides the lower transverse load degradation, is an increase of the $cos(\theta)$ correction factor leading to higher critical current density, although the effect is modest. When the wavelength is increased from 6 mm to 20 mm, the $cos(\theta)$ increases from 0.950 to 0.985, increasing the $J_c$ by 3.5 %.

Although we find improvement of the performance for lower void fraction, a warning should be given in relation to this aspect. With further reducing the void fraction of the conductor it appears that during the compaction also the length of the conductor increases [36]. This may reduce the strand diameter and at the same time decrease the $I_c$ of the conductor accordingly.

### 6.5. Recommendations

As a first test, the prediction can be verified experimentally by for example testing a short sample (in SULTAN). Two legs, identical but with a significant difference in the cabling scheme, in particular for the first four stages, can be compared directly on their performance. A cabling scheme with pitches similar, or better even some longer, than the CS2 conductor seems appropriate although eventually a more extensive parametric variation is required to examine the influence of the cabling parameters in detail. Further work is needed to analyse the optimal cabling scheme.

A test of the CSMC outer module under appropriate conditions, to explore the performance of the module with CS2 conductor having longer cabling pitches, could also serve as an indicative test.

Superc. Science & Techn. 21 April 2006.

The exact influence of the conduit axial thermal pre-compression on the critical reduction of the cable $I_c/I_{c0}(\varepsilon_b,B)$ is still unknown although experiments on strand, swaged in a steel tube may give a direction. For this part of the model we rely on a hypothesis containing some uncertainty. On this issue additional effort is required asking for a more fundamental approach looking at the 3D strain behaviour of $Nb_3Sn$ layers in relation to strand geometries.
Strands with high stiffness (Young's modulus) should be selected for conductors subjected to the highest *IxB* load if the magnet design concept allows for this choice. Strands with lower modulo can be applied in sections of the magnets with lower transverse load. In the (experimental) parametric study on cabling parameters, the influence of a lower void fraction can be further investigated, in particular in view of the strand elongation, potentially decreasing the cable cross section.
When it is confirmed that the optimisation leads to a significant minimisation of the degradation, as predicted by the model, the ITER conductor design, can be adjusted to a layout with significant economical benefit.
The use of these results can lead to a considerable reduction in the cost of superconductors for ITER. In particular when the large conductor degradation can be largely balanced. In case the time schedule does not allow for a drastic change of the conductor design, with only a change in the cabling scheme, the margin is significantly improved.

**7. Conclusions**
We developed a model that describes the design optimisation of cabled superconductors for transverse load (TEMLOP-Model) and report the first results of computations for an ITER type conductor leading to new insights, based on the measured mechanical and electromagnetic properties of an internal tin strand.
The most important conclusion of this paper is that the problem of the severe degradation of large CICCs, as designed for ITER, can be solved by increasing the cabling pitches. The principle is based on a more homogeneous distribution of the stresses and strains in the cable and significantly moderate the local peak stresses associated with short twist pitches.
The model gives a good description for the mechanical response of the strands to a transverse load, from layer to layer in the cable, based on the strand stiffness and cable void fraction. The computation is in good agreement with transverse stress-strain experiments on cables.
The simulations point out that the degradation is fully attributed to bending, although the transverse contact load on strand crossings and line contacts reaches up to 90 MPa.
We also find that a lower cable bundle void fraction and higher strand stiffness add to a further improvement of the conductor performance.

**Acknowledgements**
"This work, supported by the European Communities under the contract of Association between EURATOM/FOM, was carried out within the framework of the European Fusion Development Agreement. The views and opinions expressed herein do not necessarily reflect those of the European Commission."
The proofreading of the manuscript by my colleagues M.J. Dhalle and A. den Ouden at the University of Twente is gratefully acknowledged.